\newcommand{\be}{\begin{equation}}
\newcommand{\ee}{\end{equation}}
\newcommand{\brr}{\begin{eqnarray}}
\newcommand{\err}{\end{eqnarray}}
\newcommand{\nn}{\nonumber}
\newcommand{\bd}{\begin{displaymath}}
\newcommand{\ed}{\end{displaymath}}

\newcommand{\bib}{\bibitem}
\newcommand{\bfig}{\begin{figure}}
\newcommand{\efig}{\end{figure}}
\newcommand{\ie}{i.e.}
\newcommand{\eg}{e.g.}
\DeclareMathAlphabet{\mathpzc}{OT1}{pzc}{m}{it}
\def\alf{\alpha}
\def\bet{\beta}
\def\gam{\gamma}

\def\om{\omega}
\def\eps{\varepsilon}
\def\rpar{\right)}
\def\lpar{\left(}
\def\rbk{\right]}
\def\lbk{\left[}
\def\rbr{\right\}}
\def\lbr{\left\{}
\def\lb{\label}
\def\ro{\mbox{\boldmath $\rho$}}
\def\evo{\mbox{\boldmath $\mathcal{U}$}}
\def\dis{\mbox{\boldmath $\mathit{D}$}}
\def\fou{\mbox{\boldmath $\mathfrak{F}$}}
\def\sig{\mbox{\boldmath $\sigma$}}

\def\opp{\mbox{$\bcal{P}$}}
\def\iaf{\mbox{\scriptsize $\alpha$}}
\def\ibe{\mbox{\scriptsize $\beta$}}
\def\imu{\mbox{\scriptsize $\mu$}}
\def\inu{\mbox{\scriptsize $\nu$}}
\def\iet{\mbox{\scriptsize $\eta$}}
\def\ixi{\mbox{\scriptsize $\xi$}}
\def\iga{\mbox{\scriptsize $\gamma$}}
\def\inc{\mbox{\scriptsize ${\rm C}$}}
\def\ins{\mbox{\scriptsize ${\rm S}$}}
\def\iiet{\mbox{\tiny $\eta$}}
\def\iixi{\mbox{\tiny $\xi$}}
\def\iimu{\mbox{\tiny $\mu$}}
\def\iinu{\mbox{\tiny $\nu$}}
\def\iial{\mbox{\tiny $\alpha$}}
\def\iibt{\mbox{\tiny $\beta$}}
\def\iigm{\mbox{\tiny $\gamma$}}
\def\inn{\mbox{\tiny $\mathrm{N}$}}
\def\inh{\mbox{\tiny $\mathrm{H}$}}
\def\inq{\mbox{\tiny $Q$}}
\def\inp{\mbox{\tiny $P$}}
\def\com{\mbox{\tiny $C$}}
\def\gra{\mbox{\tiny $G$}}
\def\innu{\mbox{\tiny ${\rm U}$}}
\def\innv{\mbox{\tiny ${\rm V}$}}
\def\inno{\mbox{\tiny ${\rm O}$}}
\def\inqp{\mbox{\tiny $QP$}}
\def\rg{\rangle}
\def\lg{\langle}

\def\coloneq{\mathrel{\mathop:}=}
\def\half{\case{1}{2}}

\documentclass{iopart}
\usepackage{iopams}
\usepackage{mathrsfs}
\usepackage{amsgen}
\usepackage{amsfonts}
\usepackage{upmath}
\usepackage{upgreek}
\usepackage{yfonts}
\usepackage{dsfont}
\usepackage{bbm}
\usepackage{graphicx}
\usepackage[all]{xy}
\usepackage{ifsym}
\usepackage{pifont}
\begin{document}
\title[{\rm M A Marchiolli and M Ruzzi}]{First considerations on the generalized uncertainty principle for finite-dimensional discrete
phase spaces}
\author{M A Marchiolli and M Ruzzi}
\address{Instituto de F\'{\i}sica Te\'{o}rica, Universidade Estadual Paulista, \\
         Rua Dr Bento Teobaldo Ferraz 271, Bloco II, Barra Funda, \\
         01140-070 S\~{a}o Paulo, SP, Brazil}
\eads{\mailto{mamarchi@ift.unesp.br} and \mailto{mruzzi@ift.unesp.br}}
\begin{abstract}
Generalized uncertainty principle and breakdown of the spacetime continuum certainly represent two important results derived of various
approaches related to quantum gravity and black hole physics near the well-known Planck scale. The discreteness of space suggests, in 
particular, that all measurable lengths are quantized in units of a fundamental scale (in this case, the Planck length). Here, we propose
a self-consistent theoretical framework for an important class of physical systems characterized by a finite space of states, and show 
that such a framework enlarges previous knowledge about generalized uncertainty principles, as topological effects in finite-dimensional
discrete phase spaces come into play. Besides, we also investigate under what circumstances the generalized uncertainty principle (GUP)
works out well and its inherent limitations. 
\end{abstract}
\submitto{{\it Classical Quantum Gravity}}
%

\section{Introduction}

According to prediction of various theoretical approaches \cite{Maggiore,Kempf,Magueijo,Cortes,Alfaro,Das,Berger} associated with quantum 
gravity (for instance, String Theory and Doubly Special Relativity), as well as black hole physics, the well-known Heisenberg-Kennard-Robertson 
uncertainty principle \cite{HKR} should be replaced, near to Planck scale, by an extended counterpart --- the so-called generalized uncertainty 
principle (GUP) in current literature --- which embodies important additional terms originated from certain modified commutation relations between 
coordinate and momentum operators. The first physical implications of these particular modifications are twofold: in general, they properly reach 
not only (i) {\sl the existence of a minimum measurable length and/or a maximum measurable momentum}, but also (ii) {\sl the breakdown of the 
spacetime continuum}. Furthermore, it is worth mentioning that several quantum phenomena (such as neutrino propagation, Lamb shift, Landau levels, 
tunneling current in a scanning tunneling microscope, among others) are also affected by quantum-gravity corrections \cite{Alfaro,Das}. However, 
the discreteness of spacetime experienced at small length scales, as suggested by some theoretical frameworks that combine Quantum Mechanics and 
General Relativity \cite{Gambini}, certainly introduces important modifications in our physics concepts with profound philosophical implications 
\cite{Wharton}. Hence, spacetime continuum can now be interpreted as an emergent property whose evidence is verified only at large length scales.

In this scenario, there are different approaches to quantum gravity which introduce certain auxiliary discretization mechanisms to deal with specific
problems related to regularizations of scalar quantum field theory \cite{Johnston}, as well as dynamics and (broken) symmetries of covariant formalisms
\cite{Bianca}. Now, it is important to emphasize that all the remarkable efforts in joining two distinct theories with well-established mathematical
and physical properties only reveal the multifaceted character of the nature whose complex frontiers still remain unsolved with respect to our actual
state of knowledge. Therefore, until we figure out the intrincate mechanisms and patterns which make the nature effectively work, the theoreticians 
(in particular, the quantum-gravity community) must remain skeptical and, at the same time, open to new horizons. Recently, it has appeared in
literature an appreciable number of papers proposing similar theoretical frameworks for treating a wide class of physical systems which are
characterized by a finite-dimensional space of states --- in these descriptions, the state spaces are $N$-dimensional Hilbert spaces \cite{Vourdas}.
Besides, in connection with these finite Hilbert spaces, it should be stressed that quantum representations of $N^{2}$-dimensional discrete phase
spaces can also be constructed \cite{GP,Aldrovandi,GR1,Livine,Cotfas} and worked out in order to describe the discrete quasiprobability distribution
functions \cite{Wootters,Opatrny,Galetti1,Ruzzi2,Klimov2,Ruzzi4,Marchiolli1,Ferrie,Klimov1}. In what concerns the huge range of potential applications
associated with the discrete distribution functions, it covers different topics of particular interest in physics, such as quantum information science
\cite{Paz,Ernesto,Marchiolli2,Ferrie2}, spin-tunneling effects \cite{GR2,Marchiolli3}, open quantum systems \cite{Aolita}, and magnetic molecules
\cite{Evandro}. In this way, relevant operators whose kinematical and/or dynamical contents carry all the necessary information for describing those
physical systems are now promptly mapped in such finite-dimensional discrete phase spaces. 

The main goal of this paper is to present a self-consistent algebraic approach --- via {\sl ab initio} construction --- for finite-dimensional
discrete phase spaces that embodies convenient inherent mathematical properties which permit us to determine, from first principles, an extended
uncertainty relation for the discrete coordinate and momentum operators in the form of a GUP. The important link with the Planck units emerges
from this theoretical framework as a natural extension of certain basic quantities related to the distances between sucessive discrete eigenvalues
of $\bi{Q}$ and $\bi{P}$. Thus, the additional corrections present in the Heisenberg-Kennard-Robertson (HKR) principle include, in general, moments
of order greater than two, whose multiplicative constants are here expressed as even potencies of certain combinations involving the dimension $N$ 
of the underlying Hilbert space, the basic Planck units $\{ L_{\inp},M_{\inp} \}$, and finally, the universal physical constants $\{ \hbar,c \}$. 
The advantages and/or disadvantages of the present quantum-mechanical formulation are intrinsically connected with the inherent topology of the finite
physical system under investigation, this fact being opportunely discussed in the body of the text. Next, let us briefly mention some important points
of this particular construction process which constitute the first part of this paper. Initially formulated by Schwinger \cite{Schwinger}, the technique
of constructing unitary operator bases and the associated algebraic structure represent, in this context, two essential basic elements that lead us 
to define a mod($N$)-invariant unitary operator basis with unique characteristics: (i) {\sl it is basically constructed out by means of a unitary
transformation, via discrete displacement generator, on the parity operator} $\opp$; (ii) {\sl its mathematical properties allow to conclude that it
is a complete orthonormal operator basis}; and consequently, (iii) {\sl all the necessary quantities for describing the kinematical and dynamical
contents of a given finite physical system can now be promptly mapped upon a well-established finite-dimensional discrete phase space}. The discrete
coherent states are then formally introduced into our quantum-algebraic framework, since they represent an important example of finite quantum states
with periodic boundary conditions; in addition, it is worth stressing that certain well-known analyticity properties (namely, the non-orthogonality and
completeness relations) for the continuous counterpart \cite{Perelomov,Klauder} were properly evaluated in this case, the Jacobi theta functions
\cite{WW,Ruzzi1} playing a crucial rule in such a constructive process. 

The second part of this paper is focused basically on two important possibilities of uncertainty principles in finite-dimensional discrete phase spaces.
The first situation to be considered certainly represents an opportune moment for discussing the Robertson-Schr\"{o}dinger (RS) uncertainty principle 
\cite{Dodonov} associated with the discrete coordinate and momentum operators. To develope such a particular task, it is necessary to adapt the
previous formalism, initially conceived for unitary operators, in order to include within its scope the Hermitian coordinate and momentum operators.
Consequently, after some basic arrangements, the algebraic approach can be promptly used for evaluating, among other things, moments and mean values of
commutation and anticommutation relations involving the operators $\bi{Q}$ and $\bi{P}$ for any finite physical system where periodic boundary
conditions does not apply. It is worth emphasizing that the discussion about RS uncertainty principle and its inherent limitations open in this context
an important window of future searches for a new family of finite quantum states \cite{MOHF}. The second situation concerns a particular realization of
the unitary operators: by means of complex exponentials in which the arguments are written in terms of the discrete coordinate and momentum operators,
it has the virtues of (i) {\sl avoiding multivalued mean values}, and consequently, (ii) {\sl searching for new uncertainty relations} \cite{Massar}
(or even for new generalized entropic uncertainty relations \cite{Kraus}). In fact, the results reached in this last case pave the way for establishing
a kind of extended uncertainty principle where certain additional terms in the form of a GUP are now explicitly included. In summary, this paper is
particularly addressed to quantum-gravity community since the plethora of results here established for finite-dimensional discrete phase spaces
constitute an important (but non-conventional) theoretical approach which allows us to present a new point of view related to the GUP.   

This paper is structured as follows. In section 2, we establish certain important mathematical prerequisites to deal with unitary operators and discrete
displacement generator. In section 3, we introduce the mod($N$)-invariant unitary operator basis, as well as explore some fundamental inherent features 
in order to obtain a wide spectrum of results --- here associated with mean values, time evolution, and discrete coherent states for physical systems
described by a finite space of states  --- which constitutes our quantum-algebraic framework. Section 4 is dedicated to the uncertainty principle for
finite-dimensional discrete phase spaces where two different groups of operators are investigated: the Hermitian operators $\{ \bi{Q},\bi{P} \}$ and
a particular (but remarkable) realization of the unitary operators. Moreover, section 5 presents a possible connection with the parameters involved in 
the Planck scale and also shows how the GUP can be obtained from this algebraic approach. Finally, section 6 contains our summary and conclusions. We
have added two mathematical appendixes related to the calculational details of certain important topics and expressions used in the previous sections: 
appendix A concerns to the discrete coherent states and its inherent Wigner function (as well as the respective marginal distributions); while 
appendix B discuss the RS uncertainty principle associated with the sine and cosine operators.

\section{Prolegomenon}

In order to make the presentation of this section more self-contained, let us initially review certain essential mathematical prerequisites
related to the discrete displacement generator, for then establishing, subsequently, a set of important algebraic properties which allows us
to construct a self-consistent theoretical framework for the discrete mapping kernel $\{ \bDelta(\mu,\nu) \}_{\mu,\nu = -\ell,\ldots,\ell}$. 
This particular {\it ab initio} approach will produce the essential basic elements necessary for analysing some specific purposes associated 
with the generalized uncertainty principle.

\subsubsection*{{\bf Definition.}}
{\it Let $\{ {\bf U},{\bf  V} \}$ be a pair of unitary operators defined in a finite-dimensional state vectors space, and $\{ | u_{\alf} 
\rg, | v_{\bet} \rg \}$ the respective orthonormal eigenvectors related by the inner product $\lg u_{\alf} | v_{\bet} \rg = 
\case{1}{\sqrt{N}} \om^{\alf \bet}$ with $\om \coloneq \exp \lpar \case{2 \pi \rmi}{N} \rpar$. The general properties
\bd
\fl \quad {\bf U}^{\eta} | u_{\alf} \rg = \om^{\alf \eta} | u_{\alf} \rg , \quad {\bf V}^{\xi} | v_{\bet} \rg = \om^{\bet \xi} | 
v_{\bet} \rg , \quad {\bf U}^{\eta} | v_{\bet} \rg = | v_{\bet + \eta} \rg , \quad {\bf V}^{\xi} | u_{\alf} \rg = | u_{\alf - \xi} \rg ,
\ed
together with the fundamental relations
\bd
{\bf U}^{N} = {\bf 1} , \quad {\bf V}^{N} = {\bf 1} , \quad {\bf V}^{\xi} {\bf U}^{\eta} = \om^{\eta \xi} {\bf U}^{\eta} {\bf V}^{\xi} ,
\ed
constitute a set of basic mathematical rules that characterizes the generalized Clifford algebra} \cite{Weyl}. {\it Here, $N$ is the 
dimension of the states space and $\{ \alf,\bet,\eta,\xi \}$ corresponds to discrete labels which obey the arithmetic modulo $N$.}

A first pertinent question then emerges from our initial considerations on unitary operators and finite-dimensional state vectors: 
``Can a consistent algebraic framework for $N^{2}$-dimensional discrete phase spaces be built in order to incorporate such basic rules
and describe the main kinematical and/or dynamical features of a given physical system with a finite space of states?" To answer this
question, let us initially introduce the particular unitary operator basis \cite{Klimov1}
\be
\lb{e1}
{\bf D}(\eta,\xi) \coloneq \om^{- \{ 2^{-1} \eta \xi \}} {\bf U}^{\eta} {\bf V}^{- \xi} ,
\ee
where the labels $\eta$ and $\xi$ are associated with the dual momentum and coordinatelike variables of a $N^{2}$-dimensional discrete
phase space, and $\om^{- \{ 2^{-1} \eta \xi \}}$ consists of a specific phase whose argument obeys the following recipe: $2 \{ 2^{-1} 
\eta \xi \} \equiv \eta \xi + kN$ for all $k \in \mathbb{Z}$, namely, $2^{-1}$ denotes the operation $2^{-1} \mathrm{mod}(N)$ or also 
the multiplicative inverse of $2$ in $\mathbb{Z}_{\inn}$. Note that these labels assume integer values in the symmetric interval 
$[-\ell,\ell]$, with $\ell = \case{N-1}{2}$ fixed (henceforth, for simplicity, we assume $N$ odd throughout this paper). A comprehensive 
and useful compilation of results and properties of the aforementioned unitary operators can be found in \cite{Schwinger}, since the 
primary focus of our attention is the essential features exhibited by $\{ {\bf D}(\eta,\xi) \}_{\eta,\xi = -\ell, \ldots, \ell}$.

Next, we establish certain relevant formal properties associated with the definition proposed for ${\bf D}(\eta,\xi)$, as well as discuss
their implications on the construction process of the underlying discrete phase space \cite{Aldrovandi}.
\begin{enumerate}
\item[(i)] The inverse element ${\bf D}^{\dagger}(\eta,\xi) = {\bf D}^{-1}(\eta,\xi) = {\bf D}(-\eta,-\xi)$ exists, and consequently, the 
unitarity condition ${\bf D}^{\dagger}(\eta,\xi) {\bf D}(\eta,\xi) = {\bf D}(\eta,\xi) {\bf D}^{\dagger}(\eta,\xi) = {\bf 1}$ is totally
satisfied. In addition, the identity element is here defined through the relation ${\bf D}(0,0) = {\bf 1}$.
\item[(ii)] The auxiliary definition
\bd
\fl \quad {\bf D}(\eta,\xi) = \om^{- \{ 2^{-1} \eta \xi \}} \sum_{\gam = - \ell}^{\ell} \om^{ \gam \eta } | u_{\gam} \rg \lg u_{\gam - \xi} | 
= \om^{- \{ 2^{-1} \eta \xi \}} \sum_{\gam = - \ell}^{\ell} \om^{- \gam \xi } | v_{\gam + \eta} \rg \lg v_{\gam} |
\ed
permits us not only to calculate the trace operation
\bd
\Tr \lbk {\bf D}(\eta,\xi) \rbk = N \om^{- \{ 2^{-1} \eta \xi \}} \delta_{\eta,0}^{[\inn]} \delta_{\xi,0}^{[\inn]} \, ,
\ed
but also to find out a formal expression related to the parity operator $\opp$, \ie,
\bd
\fl \quad \opp = \frac{1}{N} \sum_{\eta,\xi = - \ell}^{\ell} {\bf D}(\eta,\xi) = \sum_{\kappa = -\ell}^{\ell} | u_{-\kappa} \rg \lg 
u_{\kappa} | = \sum_{\kappa = -\ell}^{\ell} | v_{-\kappa} \rg \lg v_{\kappa} | .
\ed
The superscript $[N]$ on the Kronecker delta denotes that this function is different from zero when its labels are congruent modulo $N$.
\item[(iii)] The action of ${\bf D}(\eta,\xi)$ on the eigenvectors $\{ | u_{\alf} \rg, | v_{\bet} \rg \}$ can be expressed as follows:
\brr
{\bf D}(\eta,\xi) | u_{\alf} \rg &=& \om^{- \{ 2^{-1} \eta \xi \} + \eta (\alf + \xi)} | u_{\alf + \xi} \rg , \nn \\
{\bf D}(\eta,\xi) | v_{\bet} \rg &=& \om^{- \{ 2^{-1} \eta \xi \} - \xi \bet} | v_{\bet + \eta} \rg . \nn
\err
These particular results show (up to phase factors) that ${\bf D}(\eta,\xi)$ generates discrete displacements in a finite mesh labelled 
by $\{ \alf,\bet \} \in [-\ell,\ell]$.
\item[(iv)] The unitarity condition verified for ${\bf D}(\eta,\xi)$ corresponds to a particular case of the composition relation
\bd
{\bf D}(\eta,\xi) {\bf D}(\eta^{\prime},\xi^{\prime}) = \om^{ \{ 2^{-1} ( \eta \xi^{\prime} - \xi \eta^{\prime} ) \} } 
{\bf D}(\eta + \eta^{\prime}, \xi + \xi^{\prime}) .
\ed
In this sense, the unitary transformation
\bd
{\bf D}(\mu,\nu) {\bf D}(\eta,\xi) {\bf D}^{\dagger}(\mu,\nu) = \om^{-( \eta \nu - \xi \mu )} {\bf D}(\eta,\xi)
\ed
represents an important element of a special set of composition relations, and it will be our main guideline in the mathematical
construction process of a unitary operator basis related to finite-dimensional discrete phase spaces.
\end{enumerate}

\section{Theoretical apparatus for the mod$(N)$-invariant unitary operator basis}

Let us introduce a formal definition for the $\mathrm{mod}(N)$-invariant unitary operator basis $\bDelta(\mu,\nu)$ through previous results 
established for the discrete displacement generator (\ref{e1}), as well as determine a set of properties which characterize its algebraic 
structure. It is worth mentioning that the phase space treated here consists of a finite mesh with $N^{2}$ points and characterized by the
discrete variables $\mu$ and $\nu$.

\subsubsection*{{\bf Definition.}}
{\it The unitary transformation ${\bf D}(\mu,\nu) \, \opp \, {\bf D}^{\dagger}(\mu,\nu)$ on the parity operator permits us to define the
mod$(N)$-invariant unitary operator basis $\{ \bDelta(\mu,\nu) \}_{\mu,\nu=-\ell,\ldots,\ell}$, whose mathematical expression is formally
written as} \cite{Klimov2}
\be
\lb{e2}
\fl \qquad \bDelta(\mu,\nu) \coloneq {\bf D}(\mu,\nu) \, \opp \, {\bf D}^{\dagger}(\mu,\nu) \equiv \frac{1}{N} \sum_{\eta,\xi = -\ell}^{\ell}
\om^{-( \eta \nu - \xi \mu )} {\bf D}(\eta,\xi) .
\ee
{\it Note that the unitary transformation on the parity operator $\opp$ and the discrete Fourier transform of the displacement generator 
${\bf D}(\eta,\xi)$ are equivalent in this case. Moreover, it is immediate to show that $\bDelta^{\dagger}(\mu,\nu) \bDelta(\mu,\nu) = 
\bDelta(\mu,\nu) \bDelta^{\dagger}(\mu,\nu) = {\bf 1}$, since $\opp^{\dagger} = \opp$ and $\opp^{2} = {\bf 1}$; consequently, the 
extra relations $\bDelta^{\dagger}(\mu,\nu) = \bDelta(\mu,\nu)$ and $\bDelta^{2}(\mu,\nu) = {\bf 1}$ can be promptly reached.}

These important particularities inherent to $\bDelta(\mu,\nu)$ can be adequately explored with the aim of establishing certain fundamental
properties which allow us not only to characterize its algebraic structure but also to conceive such operator as a discrete mapping kernel.
Thus, after some nontrivial calculations, we achieve the equalities
\brr
\fl \qquad \mathrm{(v)} \; & \frac{1}{N} \sum_{\mu,\nu = - \ell}^{\ell} \bDelta(\mu,\nu) = {\bf 1} , \nn \\
\fl \qquad \mathrm{(vi)} \; & \Tr \lbk \bDelta(\mu,\nu) \rbk = 1 , \nn \\
\fl \qquad \mathrm{(vii)} \; & \Tr \lbk \bDelta(\mu,\nu) \bDelta(\mu^{\prime},\nu^{\prime}) \rbk = N \delta_{\mu^{\prime},\mu}^{[\inn]}
\delta_{\nu^{\prime},\nu}^{[\inn]} \, , \nn \\
\fl \qquad \mathrm{(viii)} \; & \Tr \lbk \bDelta(\mu,\nu) \bDelta(\mu^{\prime},\nu^{\prime}) \bDelta(\mu^{\prime \prime},\nu^{\prime \prime}) 
\rbk = \om^{2 [ (\mu - \mu^{\prime \prime})(\nu - \nu^{\prime}) - (\mu - \mu^{\prime})(\nu - \nu^{\prime \prime}) ]} . \nn
\err
In particular, the last property has been attained with the help of the auxiliary relation
\bd
\fl \qquad \Tr \lbk {\bf D}^{\dagger}(\eta,\xi) {\bf D}(\eta^{\prime},\xi^{\prime}) {\bf D}(\eta^{\prime \prime},\xi^{\prime \prime}) 
\rbk = N \om^{ \{ 2^{-1} (\eta^{\prime} \xi^{\prime \prime} - \xi^{\prime} \eta^{\prime \prime}) \} } \delta_{\eta,\eta^{\prime} + 
\eta^{\prime \prime}}^{[\inn]} \delta_{\xi,\xi^{\prime} + \xi^{\prime \prime}}^{[\inn]} \, .
\ed
In what concerns the set of $N^{2}$ operators $\{ \bDelta(\mu,\nu) \}_{\mu,\nu=-\ell,\ldots,\ell}$, it constitutes a complete orthonormal
operator basis. This fact permits us, in principle, to construct all possible kinematical and/or dynamical quantities belonging to the 
system \cite{Aldrovandi,Schwinger}: for instance, the decomposition of any linear operator ${\bf O}$ in this basis is written as
\be
\lb{e3}
{\bf O} = \frac{1}{N} \sum_{\mu,\nu = - \ell}^{\ell} \mathscr{O}(\mu,\nu) \bDelta(\mu,\nu) ,
\ee
where the coefficients $\mathscr{O}(\mu,\nu)$ are evaluated by means of the operation $\Tr [ \bDelta(\mu,\nu) {\bf O} ]$. It must be stressed 
that this decomposition is unique since the relations (v)-(vii) are promptly verified (namely, they assert such mathematical property); in 
addition, the aforementioned coefficients correspond to a one-to-one mapping between operators and functions belonging to a finite-dimensional 
phase space characterized by the discrete variables $\mu$ and $\nu$. Henceforth, $\bDelta(\mu,\nu)$ will be recognized as a {\sl discrete mapping
kernel}. 

\subsection{Mean values}

Initially, we establish two important results related to the trace of the product of two linear operators as well as discuss their 
possible applications in quantum physics \cite{Audretsch}. The first one corresponds to the overlap
\be
\lb{e4}
\Tr [ {\bf AB} ] = \frac{1}{N} \sum_{\mu,\nu = - \ell}^{\ell} \mathscr{A}(\mu,\nu) \mathscr{B}(\mu,\nu) ,
\ee
where, in particular, the trace of the product of two density operators $\ro$ and $\sig$ coincides with the overlap of the discrete Wigner 
functions of each normalized density operator\footnote{According to equation (\ref{e3}), the decomposition of any density operator $\ro$ in 
the mod$(N)$-invariant unitary operator basis (\ref{e2}) has as coefficients the discrete Wigner function $\mathscr{W}_{\rho}(\mu,\nu) \coloneq 
\Tr [ \bDelta(\mu,\nu) \ro ]$. Pursuing this formal line of theoretical investigation, Klimov and Mu\~{n}oz \cite{Klimov2} have presented 
interesting contributions on the exact and semiclassical dynamics of $\mathscr{W}_{\rho}(\mu,\nu;t)$ for $N \gg 1$, considering the time 
evolution of certain physical systems with finite-dimensional space of states and different initial states. Their results illustrate
some potential applications of the theoretical framework here developed for $N^{2}$-dimensional phase spaces via discrete Wigner functions.},
\be
\lb{e5}
\Tr [ \ro \sig ] = \frac{1}{N} \sum_{\mu,\nu = - \ell}^{\ell} \mathscr{W}_{\rho}(\mu,\nu) \mathscr{W}_{\sigma}(\mu,\nu) .
\ee
As an immediate application of equation (\ref{e5}), let us consider the measure proposed in \cite{Paulo} of the degree of 
similarity between two quantum states $\ro$ and $\sig$, \ie, 
\be
\lb{e6}
\mathcal{F}_{\inn}(\ro,\sig) = \Tr [ \ro \sig ] + \sqrt{1 - \Tr [ \ro^{2} ]} \, \sqrt{1 - \Tr [ \sig^{2} ]} \, .
\ee
Such alternative fidelity measure is simply a sum of the Hilbert-Schmidt inner product between $\ro$ and $\sig$ and the geometric mean 
between their linear entropies; furthermore, equation (\ref{e6}) not only extends the concept of fidelity introduced by Schumacher
\cite{Schumacher} to pairs of mixed states, but also complies the Jozsa's axioms \cite{Jozsa}. Within the context of quantum information 
theory, this particular mapping between $\mathcal{F}_{\inn}(\ro,\sig)$ and discrete Wigner functions certainly represents an important
additional mathematical tool in the investigative process of entanglement for bipartite quantum systems \cite{Audretsch}.

The second result pertained to equation (\ref{e4}) is associated with the mean value
\be
\lb{e7}
\lg {\bf O} \rg \equiv \Tr [ {\bf O} \ro ] = \frac{1}{N} \sum_{\mu,\nu = - \ell}^{\ell} \mathscr{O}(\mu,\nu) \mathscr{W}_{\rho}(\mu,\nu) ,
\ee
where the evaluation of the mapped expressions $\mathscr{O}(\mu,\nu)$ and $\mathscr{W}_{\rho}(\mu,\nu)$ has a special rule in the present 
phase-space approach. To clarify this relevant point, let us initially consider two non-commuting operators ${\bf A}$ and ${\bf B}$,
as well as their inherent commutation and anticommutation relations. The mean values
\be
\lb{e8}
\lg [ {\bf A},{\bf B} ] \rg = \frac{1}{N} \sum_{\mu,\nu = - \ell}^{\ell} \bigl( [ {\bf A},{\bf B} ] \bigr) (\mu,\nu) \,
\mathscr{W}_{\rho}(\mu,\nu)
\ee
and
\be
\lb{e9}
\lg \{ {\bf A},{\bf B} \} \rg = \frac{1}{N} \sum_{\mu,\nu = - \ell}^{\ell} \bigl( \{ {\bf A},{\bf B} \} \bigr) (\mu,\nu) \,
\mathscr{W}_{\rho}(\mu,\nu)
\ee
represent two important examples of such a case, whose respective mapped expressions are conveniently written as follow:
\brr
\fl \quad \bigl( [ {\bf A},{\bf B} ] \bigr) (\mu,\nu) &=& \frac{2 \rmi}{N^{2}} \sum_{\Omega} \mathrm{Im} \lbk 
\om^{2 [ (\mu - \mu^{\prime \prime})(\nu - \nu^{\prime}) - (\mu - \mu^{\prime})(\nu - \nu^{\prime \prime}) ]} \rbk
\mathscr{A}(\mu^{\prime},\nu^{\prime}) \mathscr{B}(\mu^{\prime \prime},\nu^{\prime \prime}) , \nn \\
\fl \quad \bigl( \{ {\bf A},{\bf B} \} \bigr) (\mu,\nu) &=& \frac{2}{N^{2}} \sum_{\Omega} \mathrm{Re} \lbk 
\om^{2 [ (\mu - \mu^{\prime \prime})(\nu - \nu^{\prime}) - (\mu - \mu^{\prime})(\nu - \nu^{\prime \prime}) ]} \rbk
\mathscr{A}(\mu^{\prime},\nu^{\prime}) \mathscr{B}(\mu^{\prime \prime},\nu^{\prime \prime}) . \nn
\err
Here, $\Omega$ denotes the set $\{ \mu^{\prime},\nu^{\prime},\mu^{\prime \prime},\nu^{\prime \prime} \} \in [-\ell,\ell]$. Note that 
both expressions already show implicitly the embryonic structure of the continuous sine and cosine functions present in the well-known
Weyl-Wigner-Moyal phase space approach \cite{Leaf}. Such result is directly associated with a presymplectic structure of geometric origin
inherent to this particular finite-dimensional phase space, which is also responsible for the Poisson-like structure of the mapped commutator
expression \cite{Aldrovandi}. The next step of the present approach consists in showing some relevant aspects associated with the time
evolution of the discrete Wigner function focusing upon the existing connection between discrete phase-space propagator and Liouville function.

\subsection{Time evolution}

Following, let us now assume that $\ro(t)$ reflects the dynamics of a particular quantum system characterized by a finite space of states whose 
interaction with any dissipative environment is, in principle, automatically discarded. Consequently, if one considers only time-independent
Hamiltonians, the time evolution of $\ro(t)$ will be governed by the von Neumann-Liouville equation $\rmi \hbar \partial_{t} \ro(t) = 
[{\bf H},\ro(t)]$. In such a present approach for finite-dimensional phase spaces, the mapped equivalent of this equation describes the time 
evolution of the discrete Wigner function $\mathscr{W}_{\rho}(\mu,\nu;t) \coloneq \Tr [ \bDelta(\mu,\nu) \ro(t) ]$, \ie,
\be
\lb{e10}
\rmi \hbar \partial_{t} \mathscr{W}_{\rho}(\mu,\nu;t) = \sum_{\mu^{\prime},\nu^{\prime} = -\ell}^{\ell} \mathscr{L}(\mu,\nu,\mu^{\prime},
\nu^{\prime}) \mathscr{W}_{\rho}(\mu^{\prime},\nu^{\prime};t)
\ee
where
\bd
\fl \quad \mathscr{L}(\mu,\nu,\mu^{\prime},\nu^{\prime}) = - \frac{2 \rmi}{N^{2}} \sum_{\mu^{\prime \prime},\nu^{\prime \prime} = - \ell}^{\ell}
\mathrm{Im} \lbk \om^{2 [ (\mu - \mu^{\prime \prime})(\nu - \nu^{\prime}) - (\mu - \mu^{\prime})(\nu - \nu^{\prime \prime}) ]} \rbk
\mathscr{H}(\mu^{\prime \prime},\nu^{\prime \prime})
\ed
represents the mapped form for the Liouville operator with $\mathscr{H}(\mu^{\prime \prime},\nu^{\prime \prime})$ being equivalent to 
$\Tr [ \bDelta(\mu^{\prime \prime},\nu^{\prime \prime}) {\bf H} ]$. It is worth mentioning that equation (\ref{e10}) has a formal solution 
expressed in terms of the series \cite{GR1} 
\be
\lb{e11}
\mathscr{W}_{\rho}(\mu,\nu;t) = \sum_{\alf,\bet = -\ell}^{\ell} \mathscr{P}(\mu,\nu;t | \alf,\bet;t_{0}) \mathscr{W}_{\rho}(\alf,\bet;t_{0}) ,
\ee
whose discrete phase-space propagator
\brr
\lb{e12}
\fl \mathscr{P}(\mu,\nu;t | \alf,\bet;t_{0}) &\equiv& \frac{1}{N} \Tr \Bigl[ \bDelta(\mu,\nu) \, \evo^{\dagger} (t-t_{0}) 
\bDelta(\alf,\bet) \, \evo (t-t_{0}) \Bigr] \nn \\
&=& \delta_{\alf,\mu}^{[\inn]} \delta_{\bet,\nu}^{[\inn]} + \frac{(-\rmi)}{1! \hbar} (t-t_{0})
\mathscr{L}(\mu,\nu,\alf,\bet) \nn \\
\fl & & \!\!\!\!\! + \, \frac{(-\rmi)^{2}}{2! \hbar^{2}} (t-t_{0})^{2} \sum_{\alf^{\prime},\beta^{\prime} = - \ell}^{\ell}
\mathscr{L}(\mu,\nu,\alf^{\prime},\bet^{\prime}) \mathscr{L}(\alf^{\prime},\bet^{\prime},\alf,\bet) + \ldots
\err
permits us to evaluate directly the time evolution for the discrete Wigner function by using the series related to the iterated application 
of the mapped Liouville operator, with $\evo(t-t_{0}) \coloneq \exp \lbk - \case{\rmi}{\hbar} (t-t_{0}) {\bf H} \rbk$ denoting the associated 
time-evolution operator. The advantages and/or disadvantages from this particular mathematical solution were adequately discussed in \cite{GR2} 
for a soluble quasi-spin model\footnote{In this case, the authors used the discrete mapping kernel $\{ {\bf G}(\mu,\nu) \}_{\mu,\nu = -\ell,
\ldots,\ell}$ --- whose convenient inherent mathematical properties were formally established in \cite{Galetti1} --- with the aim of presenting, 
via time evolution of the associated discrete Wigner function, some interesting physical features intrinsic to the Lipkin-Meshkov-Glick (LMG) 
model \cite{Lipkin}. Recently, such a theoretical framework was applied with great success in exploring the spin-tunneling effects in Fe8 
magnetic clusters under the presence of external magnetic fields \cite{Evandro}.}. Some alternative solutions originated from equation (\ref{e10}) 
with different Hamiltonian operators can also be found in the current literature \cite{Klimov2,Paz}.

For instance, let us illustrate the results here obtained by means of a particular discrete counterpart of the harmonic oscillator, namely, 
the Harper Hamiltonian \cite{Harper}. This specific Hermitian operator is usually defined in terms of the Schwinger's unitary operators 
${\bf U}$ and ${\bf V}$ as follows: ${\bf H} = 2 - \half ( {\bf U} + {\bf U}^{\dagger} ) - \half ( {\bf V} + {\bf V}^{\dagger} )$. The 
corresponding mapped form $\mathscr{H}_{\inh}(\mu^{\prime \prime},\nu^{\prime \prime}) = 2 - \mathrm{Re} [ \om^{\mu^{\prime \prime}} ] - 
\mathrm{Re} [ \om^{\nu^{\prime \prime}} ]$ then leads us to find out, after some straightforward calculations, a closed-form expression for the 
Liouvillian function
\brr
\lb{e13}
\fl \mathscr{L}_{\inh}(\mu,\nu,\mu^{\prime},\nu^{\prime}) & = & 2 \rmi \mathrm{Im} [ \om^{2 (\mu \nu^{\prime} - \nu \mu^{\prime})} ] \lbk 
2 \delta_{\mu^{\prime},\mu}^{[\inn]} \delta_{\nu^{\prime},\nu}^{[\inn]} - \frac{1}{2} \lpar \delta_{\mu^{\prime},\mu + \{ 2^{-1} \} }^{[\inn]} +
\delta_{\mu^{\prime},\mu - \{ 2^{-1} \} }^{[\inn]} \rpar \delta_{\nu^{\prime},\nu}^{[\inn]} \right. \nn \\
\fl & & - \left. \frac{1}{2} \delta_{\mu^{\prime},\mu}^{[\inn]} \lpar \delta_{\nu^{\prime},\nu + \{ 2^{-1} \} }^{[\inn]} + 
\delta_{\nu^{\prime},\nu - \{ 2^{-1} \} }^{[\inn]} \rpar \rbk
\err
related to the Hamiltonian operator under scrutiny. Note that this result has a central role in our calculations since it allows not only
to determine the right-hand side of equation (\ref{e10}) but also to obtain a differential equation for the discrete phase-space propagator 
through formal solution (\ref{e11}), \ie,
\brr
\lb{e14}
\fl \rmi \hbar \partial_{t} \mathscr{P}_{\inh}(\mu,\nu;t | \alf,\bet;t_{0}) &=& - \rmi \Bigl[ \mathrm{Im} [ \om^{-\nu} ]  
\mathscr{P}_{\inh}(\mu + \{ 2^{-1} \},\nu;t | \alf,\bet;t_{0}) \nn \\
\fl & & \quad + \mathrm{Im} [ \om^{\nu} ] \mathscr{P}_{\inh}(\mu - \{ 2^{-1} \},\nu;t | \alf,\bet;t_{0}) \nn \\
\fl & & \quad + \mathrm{Im} [ \om^{\mu} ] \mathscr{P}_{\inh}(\mu,\nu + \{ 2^{-1} \};t | \alf,\bet;t_{0}) \nn \\
\fl & & \quad + \mathrm{Im} [ \om^{-\mu} ] \mathscr{P}_{\inh}(\mu,\nu - \{ 2^{-1} \};t | \alf,\bet;t_{0}) \Bigr]
\err
with the initial condition $\mathscr{P}_{\inh}(\mu,\nu;t_{0} | \alf,\bet;t_{0}) = \delta_{\alf,\mu}^{[\inn]} \delta_{\bet,\nu}^{[\inn]}$.
Consequently, the complete time-evolution of the associated discrete Wigner function will depend firstly on the exact solution of this 
differential equation and secondly, on the initial Wigner function related to $\ro(t_{0})$. In this sense, it is important to emphasize that
different mathematical methods can be applied to equation (\ref{e14}) in order to explore certain reliable numerical and/or theoretical
solutions of this particular physical model. Despite this interesting problem being relevant in the study of physical systems described by a 
finite space of states (but not crucial for present purposes), let us now introduce the discrete coherent states and discuss some inherent
properties.

\subsection{Discrete coherent states}

One of the most common states studied in quantum optics are the Klauder's coherent states \cite{Klauder} defined through the application of 
a unitary displacement operator acting on the vacuum state. This particular mathematical approach can also be extended in order to include, 
within the scope of coherent states theory \cite{Perelomov}, the discrete coherent states $| \kappa,\tau \rg \coloneq {\bf D}(\kappa,\tau) 
| 0 \rg$. The discrete vacuum state $| 0 \rg$ is here defined as being the eigenstate of the Fourier transform operator $\fou$ with eigenvalue 
equal to one, namely, this reference state obeys the eigenvalue equation $\fou | 0 \rg = 1 | 0 \rg$, whose formal solution was properly derived 
in \cite{Galetti1} (see also Ruzzi \cite{Ruzzi1} for a detailed discussion about the connection between Jacobi $\vartheta$-functions and 
discrete Fourier transforms). To illustrate such a solution, the normalized discrete wavefunction related to the reference state in the 
coordinatelike representation $\{ | u_{\gam} \rg \}_{\gam = -\ell,\ldots,\ell}$ is expressed into this context as \cite{Ruzzi2}
\be
\lb{e15}
\lg u_{\gam} | 0 \rg = \lbk \frac{2 \mathfrak{a}}{\mathscr{M}(0,0)} \rbk^{\half} \vartheta_{3} (2 \mathfrak{a} \gam | 2 \rmi \mathfrak{a}) ,
\ee
with $\mathscr{M}(0,0)$ given by\footnote{It is worth mentioning that the pattern notation for the Jacobi theta functions employed throughout 
this paper follows the Vilenkin's notation. Note that a rich material on this specific class of special functions and their respective 
properties can be found in \cite{WW}.} 
\bd
\mathscr{M}(0,0) = \sqrt{\mathfrak{a}} \bigl[ \vartheta_{3}(0| \rmi \mathfrak{a}) \vartheta_{3}(0| 4 \rmi \mathfrak{a}) + \vartheta_{4}
(0| \rmi \mathfrak{a}) \vartheta_{2}(0| 4 \rmi \mathfrak{a}) \bigr]
\ed
and $\mathfrak{a} = (2N)^{-1}$ fixed. Note that $\sqrt{2\mathfrak{a}}$ corresponds --- in equation (\ref{e15}) --- to the width of Jacobi
$\vartheta_{3}$-function and it assumes a constant value in this situation; in addition, it is important to mention that $\lg u_{\gam} | 0 \rg$ 
can also be generalized for the discrete squeezed-vaccum state \cite{Marchiolli1}. Next, we will determine a formal expression for the Wigner
function involving the discrete coherent states $\{ | \kappa,\tau \rg \}_{\kappa,\tau = -\ell,\ldots,\ell}$.

Initially, let us decompose the discrete coherent-state projetor $| \kappa,\tau \rg \lg \kappa,\tau |$ in terms of the mapping kernel
$\bDelta(\mu,\nu)$ as follows:
\be
\lb{e16}
| \kappa,\tau \rg \lg \kappa,\tau | = \frac{1}{N} \sum_{\mu,\nu = -\ell}^{\ell} \mathscr{W}_{\kappa,\tau}(\mu,\nu) \bDelta(\mu,\nu) ,
\ee
where $\mathscr{W}_{\kappa,\tau}(\mu,\nu) = \lg \kappa,\tau | \bDelta(\mu,\nu) | \kappa,\tau \rg$ represents the discrete Wigner function in 
such a case. Since $\bDelta(\mu,\nu)$ is a Hermitian operator, equation (\ref{e2}) allows us to determine an alternative expression 
for the coefficients $\mathscr{W}_{\kappa,\tau}(\mu,\nu)$ which can be interpreted, in their turn, as a specific mean value of the parity 
operator $\opp$, that is,
\be
\lb{e17}
\mathscr{W}_{\kappa,\tau}(\mu,\nu) = \lg \kappa - \mu,\tau - \nu | \opp | \kappa - \mu,\tau - \nu \rg .
\ee
After lengthy and nontrivial calculations achieved in appendix A, it is possible to show that equation (\ref{e17}) assumes the following
closed-form expression:
\be
\lb{e18}
\mathscr{W}_{\kappa,\tau}(\mu,\nu) = \mathscr{K} \lpar 2 (\kappa - \mu), 2 (\tau - \nu) \rpar .
\ee
The extra term $\mathscr{K}(2 \eta,2 \xi)$, for $\eta = \kappa - \mu$ and $\xi = \tau - \nu$ fixed, is defined in this context through the 
ratio $\mathscr{M}(2 \eta,2 \xi)/ \mathscr{M}(0,0)$, with
\bd
\fl \qquad \qquad \mathscr{M}(2 \eta,2 \xi) = \sqrt{\mathfrak{a}} \bigl[ \vartheta_{3} (2 \mathfrak{a} \eta | \rmi \mathfrak{a}) \vartheta_{3} 
(4 \mathfrak{a} \xi | 4 \rmi \mathfrak{a}) + \vartheta_{4} (2 \mathfrak{a} \eta | \rmi \mathfrak{a}) \vartheta_{2} (4 \mathfrak{a} \xi |
4 \rmi \mathfrak{a}) \bigr]
\ed
representing the function responsible for the sum of products of Jacobi theta functions evaluated at integer arguments \cite{WW}. In this
way, expansion (\ref{e16}) will lead, among other things, to generalize those results obtained in \cite{Klimov1} for the discrete vacuum 
state, as well as to investigate certain theoretical predictions originated from theories of quantum gravity for the generalized uncertainty
principle (GUP) \cite{Maggiore,Kempf,Magueijo,Cortes,Alfaro,Das,Berger}.

Following, let us now comment some few words on the inner product $\lg \kappa^{\prime},\tau^{\prime} | \kappa,\tau \rg$ between two distinct 
discrete coherent states, which is directly connected, in its turn, with the scalar product involving the discrete wavefunctions 
$\lg \kappa^{\prime},\tau^{\prime} | u_{\gam} \rg$ and $\lg u_{\gam} | \kappa,\tau \rg$ in the coordinatelike representation $\{ | u_{\gam} 
\rg \}_{\gam = -\ell,\ldots,\ell}$. Adopting the same mathematical procedure sketched out in the appendix A for the discrete Wigner function
$\mathscr{W}_{\kappa,\tau}(\mu,\nu)$, it is possible to show that such an inner product is given by
\bd
\fl \quad \lg \kappa^{\prime},\tau^{\prime} | \kappa,\tau \rg = \om^{ \{ 2^{-1} (\kappa \tau^{\prime} - \kappa^{\prime} \tau) \} }
\om^{ \half (\kappa - \kappa^{\prime}) (\tau + \tau^{\prime}) - \{ 2^{-1} (\kappa - \kappa^{\prime}) (\tau + \tau^{\prime}) \} }
\mathscr{K}(\kappa - \kappa^{\prime}, \tau - \tau^{\prime}) .
\ed
In this particular case, $\mathscr{K}(\eta,\xi)$ can be defined as $\mathscr{M}(\eta,\xi) / \mathscr{M}(0,0)$ for $\eta = \kappa -
\kappa^{\prime}$ and $\xi = \tau - \tau^{\prime}$, where the auxiliary complex function $\mathscr{M}(\eta,\xi)$ assumes the generalized
closed-form expression 
\brr
\mathscr{M}(\eta,\xi) &=& \frac{\sqrt{\mathfrak{a}}}{2} \Bigl\{ \vartheta_{3} (\mathfrak{a} \eta | \rmi \mathfrak{a}) \Bigl[
\vartheta_{3} (\mathfrak{a} \xi | \rmi \mathfrak{a}) + \rme^{\rmi \pi \eta} \vartheta_{4} (\mathfrak{a} \xi | \rmi \mathfrak{a}) \Bigr] \nn \\
& & + \rme^{\rmi \pi \xi} \vartheta_{4} (\mathfrak{a} \eta | \rmi \mathfrak{a}) \Bigl[ \vartheta_{3} (\mathfrak{a} \xi | \rmi \mathfrak{a}) -
\rme^{\rmi \pi \eta} \vartheta_{4} (\mathfrak{a} \xi | \rmi \mathfrak{a}) \Bigr] \Bigr\} . \nn
\err
It is worth mentioning that the above sum of products of Jacobi theta functions plays --- as it should be in the discrete phase space --- 
the role reserved to the Gaussian functions in the continuous counterpart. Besides, the exact expression for the overlap probability 
$| \lg \kappa^{\prime},\tau^{\prime} | \kappa,\tau \rg |^{2}$ has a direct link with the modulus square of $\mathscr{K}(\kappa - \kappa^{\prime},
\tau - \tau^{\prime})$, namely
\be
\lb{e19} 
| \lg \kappa^{\prime},\tau^{\prime} | \kappa,\tau \rg |^{2} = | \mathscr{K}(\kappa - \kappa^{\prime}, \tau - \tau^{\prime}) |^{2} ,
\ee
from which we can infer the inequality $0 < | \lg \kappa^{\prime},\tau^{\prime} | \kappa,\tau \rg |^{2} \leq 1$. For $\kappa = \kappa^{\prime}$
and $\tau = \tau^{\prime}$, this overlap probability is equal to one (normalizability condition of the inner product) and falls to zero, when
$| \kappa - \kappa^{\prime} |$ and/or $| \tau - \tau^{\prime} |$ become large enough; in other words, the discrete coherent states are not
orthogonal, as expected \cite{Perelomov}. The completeness relation in this case can be properly reached by means of the mathematical expression
\bd
\frac{1}{N} \sum_{\kappa,\tau = - \ell}^{\ell} \mathscr{W}_{\kappa,\tau}(\mu,\nu) = 1 \Longrightarrow \frac{1}{N}
\sum_{\kappa,\tau = - \ell}^{\ell} | \kappa,\tau \rg \lg \kappa,\tau | = {\bf 1} .
\ed
Note that property (v) associated with the discrete mapping kernel leads us to obtain the right-hand side of this equation.

\section{Uncertainty principle for finite-dimensional discrete phase spaces}

In the first part of this paper, we established an appreciable set of results where the kinematical and dynamical features of a given finite 
physical system can be described via discrete variables. In this second part, let us then initiate the construction process of an important
algebraic framework that allows us, in a first moment, to deal with discrete coordinate and momentum operators, through a particular realization
of the unitary operators, and its inherent uncertainty principle. The next natural step of this process consists in obtaining a self-consistent 
set of results for the unitary operators ${\bf U}$ and ${\bf V}$, culminating, in this way, with a discussion on the `uncertainty principle'
involving the respective variances $\mathscr{V}_{\innu}$ and $\mathscr{V}_{\innv}$.

\subsection{RS uncertainty principle for coordinate and momentum operators}

In this subsection, we will establish an appropriate definition of discrete coordinate and momentum operators for finite-dimensional 
spaces, which allows us to determine --- by means of the theoretical framework here described --- two important results associated with 
the mean values of their underlying commutation and anticommutation relations. In order to obtain these mean values, we basically evaluate
not only the exact mapped expressions of such commutation and anticommutation relations in finite-dimensional phase spaces labelled by the
discrete variables $\{ \mu,\nu \} \in [ -\ell,\ell ]$, but also the respective discrete Wigner function for a given density operator $\ro$.
This basic set of results then leads us to study its inherent RS uncertainty principle, as well as to discuss the different circumstances in
which such a relation does not present any inconsistency.

\subsubsection*{{\bf Definition.}}
{\it Let $\{ \bi{Q},\bi{P} \}$ be a pair of Hermitian operators defined in a $N$-dimensional state vectors space which satisfy the 
eigenvalue equations} \cite{Ruzzi3}
\bd
\bi{Q} | \mathfrak{q}_{\iaf} \rg = \mathfrak{q}_{\iaf} | \mathfrak{q}_{\iaf} \rg \qquad \mbox{and} \qquad \bi{P} | \mathfrak{p}_{\ibe} \rg = 
\mathfrak{p}_{\ibe} | \mathfrak{p}_{\ibe} \rg ,
\ed
{\it where $\mathfrak{q}_{\iaf} \coloneq \eps^{2-\delta} q_{0} \alf$ and $\mathfrak{p}_{\ibe} \coloneq \eps^{\delta} p_{0} \bet$ 
represent two distinct set of discrete eigenvalues labelled by $\{ \alf,\bet \} \in [ - \ell,\ell ]$, with $\eps = \sqrt{\case{2 \pi}{N}}$ 
fixed. Here, the free dimensionless parameter $\delta \in (0,2)$ controls the distances between successive eigenvalues of $\bi{Q}$ and 
$\bi{P}$, namely, ${\it D}_{\mathfrak{q}} = \eps^{2-\delta} q_{0}$ and ${\it D}_{\mathfrak{p}} = \eps^{\delta} p_{0}$} ({\it both the 
eigenvalue spectra are equally spaced})\footnote{Note that $\delta = 1$ corroborates the original Schwinger prescription \cite{Schwinger} 
for the unitary operators ${\bf U}$ and ${\bf V}$; on the other hand, it is important to stress that the extreme situation $\delta = 0$ 
or $2$ also recovers the quantum description of the angle-angular momentum variables in the limit $N \rightarrow \infty$ \cite{Ruzzi3}.}.
{\it Besides, the real parameters $q_{0}$ and $p_{0}$ carry units of coordinate and momentum, respectively, obeying the condition 
$q_{0} p_{0} = \hbar$ (namely, the product $q_{0} p_{0}$ coincides with the reduced Planck constant $\hbar$). Such mathematical construction 
leads us to describe the discrete coordinate and momentum operators for a finite Hilbert space.}

\subsubsection*{{\bf Remark.}}
Since $\bi{Q}$ and $\bi{P}$ are defined through the eigenstates $\{ | \mathfrak{q}_{\iaf} \rg, | \mathfrak{p}_{\ibe} \rg \}_{\iaf,\ibe =
-\ell,\ldots,\ell}$, the link with the unitary operators ${\bf U}$ and ${\bf V}$ can be promptly established:
\be
\lb{e20}
\fl \qquad {\bf U} = \exp \lpar \frac{\rmi}{\hbar} {\it D}_{\mathfrak{p}} \bi{Q} \rpar \qquad \mbox{and} \qquad {\bf V} = \exp \lpar 
\frac{\rmi}{\hbar} {\it D}_{\mathfrak{q}} \bi{P} \rpar . 
\ee
In fact, this kind of connection also reaches the discrete mapping kernel (\ref{e2}) by means of adequate changes of variables, that is,
\be
\lb{e21}
\fl \qquad \bDelta(\mathfrak{p}_{\imu},\mathfrak{q}_{\inu}) = \sum_{\mathfrak{p}_{\iiet} = - {\it R}_{\mathfrak{p}}}^{{\it R}_{\mathfrak{p}}} 
\sum_{\mathfrak{q}_{\iixi} = - {\it R}_{\mathfrak{q}}}^{{\it R}_{\mathfrak{q}}} \frac{{\it D}_{\mathfrak{p}} {\it D}_{\mathfrak{q}}}
{2 \pi \hbar} \exp \lbk \frac{\rmi}{\hbar} \lpar \mathfrak{q}_{\ixi} \mathfrak{p}_{\imu} - \mathfrak{p}_{\iet} \mathfrak{q}_{\inu} \rpar
\rbk {\bf D}(\mathfrak{p}_{\iet},\mathfrak{q}_{\ixi}) ,
\ee
where
\bd
\fl \qquad {\bf D}(\mathfrak{p}_{\iet},\mathfrak{q}_{\ixi}) = \exp \lpar - \frac{\rmi}{\hbar} \lbr 2^{-1} \mathfrak{p}_{\iet} 
\mathfrak{q}_{\ixi} \rbr \rpar \exp \lpar \frac{\rmi}{\hbar} \mathfrak{p}_{\iet} \bi{Q} \rpar \exp \lpar - \frac{\rmi}{\hbar} 
\mathfrak{q}_{\ixi} \bi{P} \rpar 
\ed
represents the discrete displacement generator (\ref{e1}) written in terms of $\mathfrak{p}_{\iet} = {\it D}_{\mathfrak{p}} \eta$ and
$\mathfrak{q}_{\ixi} = {\it D}_{\mathfrak{q}} \xi$. The limits ${\it R}_{\mathfrak{p}} = \ell {\it D}_{\mathfrak{p}}$ and
${\it R}_{\mathfrak{q}} = \ell {\it D}_{\mathfrak{q}}$ appeared in the finite double series correspond to the maximum range of each 
discrete spectrum $\{ \mathfrak{p}_{\iet} \}$ and $\{ \mathfrak{q}_{\ixi} \}$. So, if one considers the continuum limit $N \rightarrow 
\infty$ in such a case, we will reobtain the well-known Cahill-Glauber formalism \cite{Glauber} through the continuous mapping kernel
$\bDelta(p,q)$.

The closed-form relations for $N^{2}$-dimensional phase-space representatives of the discrete coordinate and momentum operators are 
evaluated in this context by means of the specific trace operations
\bd
\fl \quad \bigl( \bi{Q} \bigr) (\mathfrak{p}_{\imu},\mathfrak{q}_{\inu}) \equiv \Tr [ \bDelta(\mathfrak{p}_{\imu},\mathfrak{q}_{\inu})
\bi{Q} ] = \mathfrak{q}_{\inu} \quad \mbox{and} \quad \bigl( \bi{P} \bigr) (\mathfrak{p}_{\imu},\mathfrak{q}_{\inu}) \equiv \Tr [
\bDelta(\mathfrak{p}_{\imu},\mathfrak{q}_{\inu}) \bi{P} ] = \mathfrak{p}_{\imu} .
\ed
Let us now apply such a mathematical recipe in order to obtain equivalent expressions for the commutation and anticommutation 
relations. Thus, after some straightforward but tedious calculations, we achieve the algebraic expressions
\be
\lb{e22}
\fl \bigl( [ \bi{Q},\bi{P} ] \bigr) (\mathfrak{p}_{\imu},\mathfrak{q}_{\inu}) = 2 \rmi \sum_{\mathfrak{p}_{\iiet} = - 
{\it R}_{\mathfrak{p}}}^{{\it R}_{\mathfrak{p}}} \sum_{\mathfrak{q}_{\iixi} = - {\it R}_{\mathfrak{q}}}^{{\it R}_{\mathfrak{q}}} 
\frac{{\it D}_{\mathfrak{p}} {\it D}_{\mathfrak{q}}}{2 \pi \hbar} \mathfrak{p}_{\iiet} \mathfrak{q}_{\iixi} \sin \lbk \frac{2}{\hbar}
( \mathfrak{p}_{\iiet} - \mathfrak{p}_{\imu} ) ( \mathfrak{q}_{\iixi} - \mathfrak{q}_{\inu} ) \rbk
\ee
and
\be
\lb{e23}
\fl \bigl( \{ \bi{Q},\bi{P} \} \bigr) (\mathfrak{p}_{\imu},\mathfrak{q}_{\inu}) = 2 \sum_{\mathfrak{p}_{\iiet} = - 
{\it R}_{\mathfrak{p}}}^{{\it R}_{\mathfrak{p}}} \sum_{\mathfrak{q}_{\iixi} = - {\it R}_{\mathfrak{q}}}^{{\it R}_{\mathfrak{q}}} 
\frac{{\it D}_{\mathfrak{p}} {\it D}_{\mathfrak{q}}}{2 \pi \hbar} \mathfrak{p}_{\iiet} \mathfrak{q}_{\iixi} \cos \lbk \frac{2}{\hbar}
( \mathfrak{p}_{\iiet} - \mathfrak{p}_{\imu} ) ( \mathfrak{q}_{\iixi} - \mathfrak{q}_{\inu} ) \rbk .
\ee
In fact, equations (\ref{e22}) and (\ref{e23}) represent two basic ingredients within the qualitative and/or quantitative evaluations of 
the mean values
\brr
\lb{e24}
\fl \qquad \lg [ \bi{Q},\bi{P} ] \rg &=& \sum_{\mathfrak{p}_{\iimu} = - {\it R}_{\mathfrak{p}}}^{{\it R}_{\mathfrak{p}}} 
\sum_{\mathfrak{q}_{\iinu} = - {\it R}_{\mathfrak{q}}}^{{\it R}_{\mathfrak{q}}} \frac{{\it D}_{\mathfrak{p}} {\it D}_{\mathfrak{q}}}
{2 \pi \hbar} \bigl( [ \bi{Q},\bi{P} ] \bigr) (\mathfrak{p}_{\imu},\mathfrak{q}_{\inu}) \, 
\mathscr{W}_{\rho}(\mathfrak{p}_{\imu},\mathfrak{q}_{\inu}) , \\
\lb{e25}
\fl \qquad \lg \{ \bi{Q},\bi{P} \} \rg &=& \sum_{\mathfrak{p}_{\iimu} = - {\it R}_{\mathfrak{p}}}^{{\it R}_{\mathfrak{p}}} 
\sum_{\mathfrak{q}_{\iinu} = - {\it R}_{\mathfrak{q}}}^{{\it R}_{\mathfrak{q}}} \frac{{\it D}_{\mathfrak{p}} {\it D}_{\mathfrak{q}}}
{2 \pi \hbar} \bigl( \{ \bi{Q},\bi{P} \} \bigr) (\mathfrak{p}_{\imu},\mathfrak{q}_{\inu}) \, 
\mathscr{W}_{\rho}(\mathfrak{p}_{\imu},\mathfrak{q}_{\inu}) .
\err
Another important ingredient refers to the initial quantum state $\ro$ associated with a physical system belonging to a finite-dimensional 
space of states and here described by means of the discrete Wigner function $\mathscr{W}_{\rho}(\mathfrak{p}_{\imu},\mathfrak{q}_{\inu})$, 
that is, the kinematical and/or dynamical contents of such a system can now be promptly mapped on $N^{2}$-dimensional phase spaces via
quasiprobability distribution function $\mathscr{W}_{\rho}(\mathfrak{p}_{\imu},\mathfrak{q}_{\inu})$. Hence, expressions (\ref{e24}) and 
(\ref{e25}) carry all relevant quantum correlations inherent to the system of interest, and this fact will be carefully explored in our 
investigation on uncertainty principles.

Following, let us establish two complementary properties for the discrete mapping kernel $\bDelta(\mathfrak{p}_{\imu},\mathfrak{q}_{\inu})$,
as well as to discuss their immediate applications on the evaluation process of the mean values $\lg \bi{Q}^{k} \rg$ and $\lg \bi{P}^{k} \rg$
for $k \in \mathbb{N}$. Such particular properties are reached through the individual sums on the whole discrete spectrum 
$\{ \mathfrak{p}_{\imu} \}$ and $\{ \mathfrak{q}_{\inu} \}$, namely,
\be
\lb{e26}
\fl \;\; \sum_{\mathfrak{p}_{\iimu} = - {\it R}_{\mathfrak{p}}}^{{\it R}_{\mathfrak{p}}} \frac{{\it D}_{\mathfrak{p}}}{2 \pi \hbar}
\bDelta(\mathfrak{p}_{\imu},\mathfrak{q}_{\inu}) = | \mathfrak{q}_{\inu} \rg \lg \mathfrak{q}_{\inu} | \quad \mbox{and} \quad \!\!
\sum_{\mathfrak{q}_{\iinu} = - {\it R}_{\mathfrak{q}}}^{{\it R}_{\mathfrak{q}}} \frac{{\it D}_{\mathfrak{q}}}{2 \pi \hbar}
\bDelta(\mathfrak{p}_{\imu},\mathfrak{q}_{\inu}) = | \mathfrak{p}_{\imu} \rg \lg \mathfrak{p}_{\imu} | .
\ee
It is worth mentioning that $\{ | \mathfrak{p}_{\imu} \rg, | \mathfrak{q}_{\inu} \rg \}$ are connected with the eigenvectors $\{ | v_{\imu} 
\rg, | u_{\inu} \rg \}$ through the scaling relations $| \mathfrak{p}_{\imu} \rg = {\it D}_{\mathfrak{p}}^{-1/2} | v_{\imu} \rg$ and 
$| \mathfrak{q}_{\inu} \rg = {\it D}_{\mathfrak{q}}^{-1/2} | u_{\inu} \rg$. As a byproduct of the identities (\ref{e26}), the marginal 
distributions related to discrete Wigner function can now be properly defined as follow:
\brr
\mathscr{Q}_{\rho}(\mathfrak{q}_{\inu}) &=& \sum_{\mathfrak{p}_{\iimu} = - {\it R}_{\mathfrak{p}}}^{{\it R}_{\mathfrak{p}}}
\frac{{\it D}_{\mathfrak{p}}}{2 \pi \hbar} \mathscr{W}_{\rho}(\mathfrak{p}_{\imu},\mathfrak{q}_{\inu}) = \lg \mathfrak{q}_{\inu} | \ro |
\mathfrak{q}_{\inu} \rg , \nn \\
\mathscr{R}_{\rho}(\mathfrak{p}_{\imu}) &=& \sum_{\mathfrak{q}_{\iinu} = - {\it R}_{\mathfrak{q}}}^{{\it R}_{\mathfrak{q}}}
\frac{{\it D}_{\mathfrak{q}}}{2 \pi \hbar} \mathscr{W}_{\rho}(\mathfrak{p}_{\imu},\mathfrak{q}_{\inu}) = \lg \mathfrak{p}_{\imu} | \ro |
\mathfrak{q}_{\imu} \rg . \nn
\err
These marginal distributions have an important role in the evaluations of the mean values of moments associated with the discrete coordinate
and momentum operators, that is
\be
\lb{e27}
\fl \qquad \lg \bi{Q}^{k} \rg = \sum_{\mathfrak{q}_{\iinu} = - {\it R}_{\mathfrak{q}}}^{{\it R}_{\mathfrak{q}}} {\it D}_{\mathfrak{q}} \,
\mathfrak{q}_{\inu}^{k} \, \mathscr{Q}_{\rho}(\mathfrak{q}_{\inu}) \quad \mbox{and} \quad \lg \bi{P}^{k} \rg = \sum_{\mathfrak{p}_{\iimu} = - 
{\it R}_{\mathfrak{p}}}^{{\it R}_{\mathfrak{p}}} {\it D}_{\mathfrak{p}} \, \mathfrak{p}_{\imu}^{k} \, \mathscr{R}_{\rho}(\mathfrak{p}_{\imu}) ,
\ee
since $\mathscr{Q}_{\rho}(\mathfrak{q}_{\inu})$ and $\mathscr{R}_{\rho}(\mathfrak{p}_{\imu})$ can be considered within this context as 
statistical weights for each particular situation. As an immediate application of these results, let us now investigate qualitatively the
RS uncertainty principle \cite{Dodonov}.

Initially, let us introduce two well-known expressions for the variances related to the discrete position and momentum operators, namely,
\bd
\mathscr{V}_{\inq} \equiv \lg \bi{Q}^{2} \rg - \lg \bi{Q} \rg^{2} \qquad \mbox{and} \qquad \mathscr{V}_{\inp} \equiv \lg \bi{P}^{2} \rg - 
\lg \bi{P} \rg^{2} ,
\ed
as well as the covariance function $\mathscr{C}_{\inqp} \equiv \lg \half \{ \bi{Q},\bi{P} \} \rg - \lg \bi{Q} \rg \lg \bi{P} \rg$ which
presents a direct link with the anticommutation relation mean value. These specific equations allow us, in particular, to define the RS 
uncertainty relation as follows \cite{Dodonov}:
\be
\lb{e28}
\mathscr{U}_{\inqp} \coloneq \mathscr{V}_{\inq} \mathscr{V}_{\inp} - \lpar \mathscr{C}_{\inqp} \rpar^{2} \geq \case{1}{4} \left| \lg
[ \bi{Q},\bi{P} ] \rg \right|^{2} .
\ee
Next, let us consider a physical system $\ro$ defined in a finite-dimensional space of states where periodic boundary conditions does not
apply. Since 
\bd
\bigl\{ \bDelta(\mathfrak{p}_{\imu},\mathfrak{q}_{\inu}) : \mathfrak{p}_{\imu} \in [ -{\it R}_{\mathfrak{p}},{\it R}_{\mathfrak{p}} ] 
\;\; \mbox{and} \;\; \mathfrak{q}_{\inu} \in [ -{\it R}_{\mathfrak{q}},{\it R}_{\mathfrak{q}} ] \bigr\} 
\ed
represents a complete orthonormal operator basis, the discrete Wigner function and its respective marginal distributions do not present 
into this theoretical approach any tangible mathematical inconsistencies associated with the mapping of $\ro$ upon a $N^{2}$-dimensional 
discrete phase space. Therefore, the variances $\mathscr{V}_{\inq}$ and $\mathscr{V}_{\inp}$ are single-valued functions, and consequently, 
equation (\ref{e28}) yields a well-defined inequality. Now, let $\ro$ describe a physical system with well-established periodic boundary
conditions. In this specific case, the mean values appeared in (\ref{e28}) are multivalued and such an apparent inconsistency is hard to
be worked out. To overcome this particular problem, it turns natural to define unitary operators as complex exponential functions whose
arguments are expressed in terms of those discrete coordinate and momentum operators --- see equation (\ref{e20}). Such a mathematical
procedure avoids, on the one hand, certain problems involving multivalued mean values and, on the other hand, it shows an alternative way
of looking for new uncertainty principles where an important class of physical systems can now be adequately studied \cite{Phase}.

\subsection{Uncertainty principle for unitary operators}

After constructing a solid theoretical framework for finite-dimensional discrete phase spaces, let us now implement a self-consistent set
of mathematical and physical results involving the unitary operators given in Eq. (\ref{e20}). In order to realize this purpose, we 
initially verify that ${\bf U}$ and ${\bf V}$ can also be described by particular discrete displacement generators --- namely, ${\bf U} = 
{\bf D}({\it D}_{\mathfrak{p}},0)$ and ${\bf V} = {\bf D}^{\dagger}(0,{\it D}_{\mathfrak{q}})$ --- for then establishing, as a second and
important intermediate step in our calculations, their respective variances through the well-known mathematical expression $\mathscr{V}_{\inno}
= \lg \delta {\bf O} \delta {\bf O}^{\dagger} \rg$ with $\delta {\bf O} \coloneq {\bf O} - \lg {\bf O} \rg$. It is worth stressing that
the results here reported corroborate and generalize those obtained by Massar and Spindel \cite{Massar} in connection with uncertainty
principle derived for the discrete Fourier transform.

\subsubsection*{{\bf Definition.}}
{\it Let $\mathscr{V}_{\innu} \coloneq 1 - | \lg {\bf U} \rg |^{2}$ and $\mathscr{V}_{\innv} \coloneq 1 - | \lg {\bf V} \rg |^{2}$ 
characterize, respectively, the variances related to the unitary operators ${\bf U} = {\bf D}({\it D}_{\mathfrak{p}},0)$ and ${\bf V} = 
{\bf D}^{\dagger}(0,{\it D}_{\mathfrak{q}})$, where $\lg {\bf U} \rg$ and $\lg {\bf V} \rg$ represent two mean values defined in a
$N$-dimensional state vectors space as follow:}
\brr
\fl \quad \lg {\bf U} \rg \coloneq \Tr \lbk {\bf D}({\it D}_{\mathfrak{p}},0) \ro \rbk = \sum_{\mathfrak{q}_{\iinu} = - 
{\it R}_{\mathfrak{q}}}^{{\it R}_{\mathfrak{q}}} {\it D}_{\mathfrak{q}} \exp \lpar \frac{\rmi}{\hbar} {\it D}_{\mathfrak{p}} 
\mathfrak{q}_{\inu} \rpar \mathscr{Q}_{\rho}(\mathfrak{q}_{\inu}) = \sum_{k=0}^{\infty} \frac{( \rmi {\it D}_{\mathfrak{p}} )^{k}}
{\hbar^{k} k!} \lg \bi{Q}^{k} \rg , \nn \\
\fl \quad \lg {\bf V} \rg \coloneq \Tr \lbk {\bf D}^{\dagger}(0,{\it D}_{\mathfrak{q}}) \ro \rbk = \sum_{\mathfrak{p}_{\iimu} = - 
{\it R}_{\mathfrak{p}}}^{{\it R}_{\mathfrak{p}}} {\it D}_{\mathfrak{p}} \exp \lpar \frac{\rmi}{\hbar} {\it D}_{\mathfrak{q}} 
\mathfrak{p}_{\imu} \rpar \mathscr{R}_{\rho}(\mathfrak{p}_{\imu}) = \sum_{k=0}^{\infty} \frac{( \rmi {\it D}_{\mathfrak{q}} )^{k}}
{\hbar^{k} k!} \lg \bi{P}^{k} \rg . \nn
\err
{\it Since $\ro$ refers to a normalized density operator, the Cauchy-Schwarz inequality leads us to prove that $| \lg {\bf U} \rg |^{2}$ and 
$| \lg {\bf V} \rg |^{2}$ are restricted to the closed interval $[0,1]$; consequently, both the variances are trivially bounded by $0 \leq
\mathscr{V}_{\innu} \leq 1$ and $0 \leq \mathscr{V}_{\innv} \leq 1$} \cite{Jauch}. {\it In fact, the upper and lower bounds are promptly reached 
when one considers the localized bases $\lbr | \bar{\mathfrak{q}}_{\inu} \rg \lg \bar{\mathfrak{q}}_{\inu} | \rbr_{-{\it R}_{\mathfrak{q}} 
\leq \bar{\mathfrak{q}}_{\iinu} \leq {\it R}_{\mathfrak{q}}}$ and $\lbr | \bar{\mathfrak{p}}_{\imu} \rg \lg \bar{\mathfrak{p}}_{\imu} | 
\rbr_{-{\it R}_{\mathfrak{p}} \leq \bar{\mathfrak{p}}_{\iimu} \leq {\it R}_{\mathfrak{p}}}$, \ie, for $\ro = | \bar{\mathfrak{q}}_{\inu} \rg 
\lg \bar{\mathfrak{q}}_{\inu} | \Rightarrow \mathscr{V}_{\innu} = 0$ and $\mathscr{V}_{\innv} = 1$; otherwise, if $\ro = | \bar{\mathfrak{p}}_{\imu}
\rg \lg \bar{\mathfrak{p}}_{\imu} | \Rightarrow \mathscr{V}_{\innu} = 1$ and $\mathscr{V}_{\innv} = 0$. Besides, $\mathscr{V}_{\innu}$ and
$\mathscr{V}_{\innv}$ do not change under the phase transformation ${\bf U} \rightarrow e^{\rmi \varphi} {\bf U}$ and ${\bf V} \rightarrow
e^{\rmi \theta} {\bf V}$.}

\subsubsection*{{\bf Remark 1.}}
The discrete coherent states represent an intermediate situation for the aforementioned variances. Such statement can be immediately 
demonstrated through the analytical expressions
\brr
\fl \quad \mathscr{V}_{\innu} = 1 - | \lg \mathfrak{p}_{\kappa},\mathfrak{q}_{\tau} | {\bf D}({\it D}_{\mathfrak{p}},0) | 
\mathfrak{p}_{\kappa},\mathfrak{q}_{\tau} \rg |^{2} = 1 - | \lg 0,0 | {\it D}_{\mathfrak{p}},0 \rg |^{2} = 1 - \mathscr{K}^{2}
({\it D}_{\mathfrak{p}},0) , \nn \\
\fl \quad \mathscr{V}_{\innv} = 1 - | \lg \mathfrak{p}_{\kappa},\mathfrak{q}_{\tau} | {\bf D}^{\dagger}(0,{\it D}_{\mathfrak{q}}) | 
\mathfrak{p}_{\kappa},\mathfrak{q}_{\tau} \rg |^{2} = 1 - | \lg 0,0 | 0,-{\it D}_{\mathfrak{q}} \rg |^{2} = 1 - \mathscr{K}^{2}
(0,-{\it D}_{\mathfrak{q}}) . \nn
\err
Note that $\mathscr{K}({\it D}_{\mathfrak{p}},0)$ and $\mathscr{K}(0,-{\it D}_{\mathfrak{q}})$ are formally equivalent to $\mathscr{K}(1,0)$
and $\mathscr{K}(0,-1)$, respectively; furthermore, due to the mathematical properties inherent to Jacobi theta functions, it is easy to 
show that $\mathscr{K}(1,0) \equiv \mathscr{K}(0,-1)$ which implies in $\mathscr{V}_{\innu} = \mathscr{V}_{\innv}$. This particular result 
reinforces the `minimum uncertainty states character' related to the finite-dimensional discrete coherent states here studied. As an  
illustration of this case, Figure 1 shows the plot of $\mathscr{V}_{\innu (\innv)}$ as a function of the state vectors space dimension
$N$, where certain peculiarities deserve be mentioned: the first one consists in noticing that $\mathscr{V}_{\innu (\innv)} > \case{1}{2}$
for $N=3$, confirming, in this way, specific results obtained by Opatrn\'{y} {\it et al} \cite{Opatrny} in the study of number-phase
uncertainty relations; the second one is related to the asymptotic limit $\mathscr{V}_{\innu (\innv)} \geq \case{\pi}{N}$ when $N \gg 1$,
which also corroborates a group of important partial results derived in \cite{Massar}. 
\begin{figure}[!t]
\centering
\begin{minipage}[b]{0.6\linewidth}
\includegraphics[width=\textwidth]{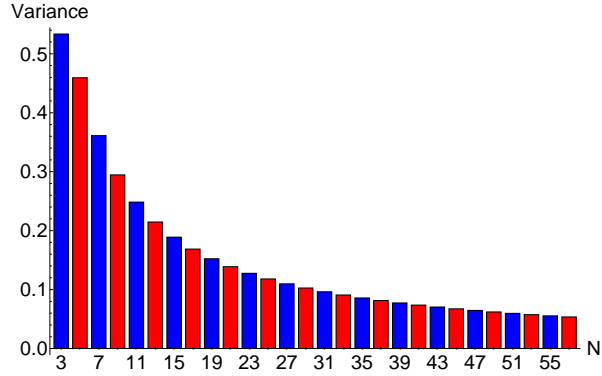}
\end{minipage}
\caption{Plot of $\mathscr{V}_{\innu (\innv)}$ versus $N \in [3,57]$ for the discrete coherent states defined in equation (\ref{e16}), where 
$N$ is associated with the state vectors space dimension. In such a case, the asymptotic behaviour verified for $\mathscr{V}_{\innu (\innv)}$
when $N \gg 1$ satisfies numerically the mathematical relation $\mathscr{V}_{\innu (\innv)} \geq \case{\pi}{N}$.} 
\end{figure}

In accordance with the algebraic approach developed until now, let us determine in this moment an important inequality for the unitary 
operators through the results derived in appendix B involving the sine and cosine operators. For this particular task, let us initially
introduce a set of preliminary propositions that leads us to establish an indirect mathematical proof of the main result. Thus, the first 
one consists in obtaining an upper bound for the product of certain variances related to $\lbr {\bf C}_{\innu},{\bf S}_{\innu},{\bf C}_{\innv},
{\bf S}_{\innv} \rbr$ with the help of $\mathscr{V}_{\innu}$ and $\mathscr{V}_{\innv}$. In fact, this direct result shows simply that each
pair $\mathscr{V}_{\inc_{\innu}} \mathscr{V}_{\inc_{\innv}}$, $\mathscr{V}_{\inc_{\innu}} \mathscr{V}_{\ins_{\innv}}$,
$\mathscr{V}_{\ins_{\innu}} \mathscr{V}_{\inc_{\innv}}$, and $\mathscr{V}_{\ins_{\innu}} \mathscr{V}_{\ins_{\innv}}$ has as superior limit
the product $\mathscr{V}_{\innu} \mathscr{V}_{\innv}$.

\subsubsection*{{\bf Proposition 1.}} 
The inequalities $\mathscr{V}_{\innu} \mathscr{V}_{\innv} \geq \lbr \mathscr{V}_{\inc_{\innu}} \mathscr{V}_{\inc_{\innv}},
\mathscr{V}_{\inc_{\innu}} \mathscr{V}_{\ins_{\innv}},\mathscr{V}_{\ins_{\innu}} \mathscr{V}_{\inc_{\innv}}, \mathscr{V}_{\ins_{\innu}}
\mathscr{V}_{\ins_{\innv}} \rbr$ hold for any density operator $\ro$ belonging to a finite-dimensional state vectors space.

\subsubsection*{{\bf Proof.}}
To begin with, note that $\mathscr{V}_{\innu} \mathscr{V}_{\innv}$ can be expressed as $\lpar \mathscr{V}_{\inc_{\innu}} + 
\mathscr{V}_{\ins_{\innu}} \rpar \lpar \mathscr{V}_{\inc_{\innv}} + \mathscr{V}_{\ins_{\innv}} \rpar$ for any quantum state $\ro$ properly
defined in a finite-dimensional state vectors space; consequently, the inequalities mentioned previously follow immediately from a simple
inspection of such mathematical relation.

The second one establishes a trivial but important result showing the connection between $\mathscr{V}_{\innu(\innv)}$ and the quadratic mean 
values related to the sine and cosine operators, that is, $\mathscr{V}_{\innu(\innv)} = 1 - \lpar \lg {\bf C}_{\innu(\innv)} \rg^{2} + \lg
{\bf S}_{\innu(\innv)} \rg^{2} \rpar$. It is worth stressing that the mathematical procedure here used to demonstrate such a relation presents
certain essential basic elements very useful to our understanding of future discussions on unitary operators and generalized uncertainty
principle, as well as its inherent link.

\subsubsection*{{\bf Proposition 2.}}
The sums $\lg {\bf C}_{\innu} \rg^{2} + \lg {\bf S}_{\innu} \rg^{2}$ and $\lg {\bf C}_{\innv} \rg^{2} + \lg {\bf S}_{\innv} \rg^{2}$
involving the quadratic mean values of the sine and cosine operators are directly connected with the quadratic absolute values 
$| \lg {\bf U} \rg |^{2}$ and $| \lg {\bf V} \rg |^{2}$, respectively.

\subsubsection*{{\bf Proof.}}
As a first step in our proof, let us decompose the normalized density operator in diagonal and non-diagonal parts for both the discrete 
momentum and coordinatelike representations. This particular mathematical form permits us to write the actions of the cosine operators on 
$\ro$ as follow:
\brr
\fl \; \; {\bf C}_{\innu} \ro &=& \sum_{\mathfrak{q}_{\iial} \in \lbk - {\it R}_{\mathfrak{q}},{\it R}_{\mathfrak{q}} \rbk} | c_{\iaf} |^{2} 
\cos \lpar \frac{{\it D}_{\mathfrak{p}} \mathfrak{q}_{\iaf}}{\hbar} \rpar | \mathfrak{q}_{\iaf} \rg \lg \mathfrak{q}_{\iaf} | \nn \\
\fl & & + \sum_{{\mathfrak{q}_{\iial},\mathfrak{q}_{\iibt} \in \lbk - {\it R}_{\mathfrak{q}},{\it R}_{\mathfrak{q}} \rbk} \atop
{\mathfrak{q}_{\iial} \neq \mathfrak{q}_{\iibt}}} c_{\iaf} c_{\ibe}^{\ast} \exp \lbk \frac{\rmi {\it D}_{\mathfrak{p}}}{2 \hbar} \lpar
\mathfrak{q}_{\iaf} - \mathfrak{q}_{\ibe} \rpar \rbk \cos \lbk \frac{{\it D}_{\mathfrak{p}}}{2 \hbar} \lpar \mathfrak{q}_{\iaf} + 
\mathfrak{q}_{\ibe} \rpar \rbk | \mathfrak{q}_{\iaf} \rg \lg \mathfrak{q}_{\ibe} | \nn
\err
and
\brr
\fl \; \; {\bf C}_{\innv} \ro &=& \sum_{\mathfrak{p}_{\iiet} \in \lbk - {\it R}_{\mathfrak{p}},{\it R}_{\mathfrak{p}} \rbk} | \tilde{c}_{\iet} |^{2}
\cos \lpar \frac{{\it D}_{\mathfrak{q}} \mathfrak{p}_{\iet}}{\hbar} \rpar | \mathfrak{p}_{\iet} \rg \lg \mathfrak{p}_{\iet} | \nn \\
\fl & & + \sum_{{\mathfrak{p}_{\iiet},\mathfrak{p}_{\iigm} \in \lbk - {\it R}_{\mathfrak{p}},{\it R}_{\mathfrak{p}} \rbk} \atop 
{\mathfrak{p}_{\iiet} \neq \mathfrak{p}_{\iigm}}} \tilde{c}_{\iet} \tilde{c}_{\iga}^{\ast} \exp \lbk \frac{\rmi {\it D}_{\mathfrak{q}}}{2 \hbar} 
\lpar \mathfrak{p}_{\iiet} - \mathfrak{p}_{\iigm} \rpar \rbk \cos \lbk \frac{{\it D}_{\mathfrak{q}}}{2 \hbar} \lpar \mathfrak{p}_{\iet} +
\mathfrak{p}_{\iga} \rpar \rbk | \mathfrak{p}_{\iet} \rg \lg \mathfrak{p}_{\iga} | . \nn
\err
Similar procedure can also be applied for the sine operators ${\bf S}_{\innu}$ and ${\bf S}_{\innv}$. The coefficients $c_{\iaf(\ibe)} 
\equiv \lg \mathfrak{q}_{\iaf(\ibe)} | \Psi \rg$ and $\tilde{c}_{\iet(\iga)} \equiv \lg \mathfrak{p}_{\iet(\iga)} | \Psi \rg$ denote, in 
this situation, the different discrete representations of the wavefunction $| \Psi \rg$; moreover, both the coefficients are connected 
through a discrete Fourier transform. Following, it is interesting to note that the trace operation eliminates the non-diagonal parts of 
these decompositions, the remaining diagonal parts now being responsible for the effective contributions related to the mean values
$\lg {\bf C}_{\innu(\innv)} \rg$ and $\lg {\bf S}_{\innu(\innv)} \rg$. Consequently, since the mean values $\lg {\bf U} \rg$ and
$\lg {\bf V} \rg$ can be written as $\lg {\bf U} \rg = \lg {\bf C}_{\innu} \rg + \rmi \lg {\bf S}_{\innu} \rg$ and $\lg {\bf V} \rg = 
\lg {\bf C}_{\innv} \rg + \rmi \lg {\bf S}_{\innv} \rg$, it turns out to be immediate to obtain the desired results.

The main result of this section is based on the Wiener-Kinchin theorem for signal processing and provides a constraint between the values
of $\lg {\bf V}^{\xi} \rg$ (correlation function) and $\lg {\bf U}^{\eta} \rg$ (discrete Fourier transform of the intensity time series).
According to Massar and Spindel \cite{Massar}: `This kind of constraint should prove useful in signal processing, as it constrains what
kinds of signals are possible, or what kind of wavelet bases one can construct.' Next, let us establish this result by means of the theorem 
below, for then proceeding with a numerical study of such a theoretical statement and its implications. 

\subsubsection*{{\bf Theorem ({\sl Massar and Spindel}).}}
{\it Let ${\bf U}$ and ${\bf V}$ be two unitary operators which obey the generalized Clifford algebra within a $N$-dimensional state 
vectors space for $N \geq 2$. The variances $\mathscr{V}_{\innu} \coloneq 1 - | \lg {\bf U} \rg |^{2}$ and $\mathscr{V}_{\innv} \coloneq 
1 - | \lg {\bf V} \rg |^{2}$, here defined for a given quantum state $\ro$ and limited to the closed interval $\mathscr{V}_{\innu(\innv)}
\in [0,1]$, allow us to establish the bound
\be
\lb{e29}
\mathscr{U}_{\innu \innv} \coloneq (1+2A) \mathscr{V}_{\innu} \mathscr{V}_{\innv} + A^{2} (\mathscr{V}_{\innu} + \mathscr{V}_{\innv} - 1 ) 
\geq 0
\ee
with $A = \tan \lpar \case{\pi}{N} \rpar$ and $0 \leq A < \infty$. The saturation is reached for localized bases.}

\begin{figure}[!t]
\centering
\begin{minipage}[b]{0.6\linewidth}
\includegraphics[width=\textwidth]{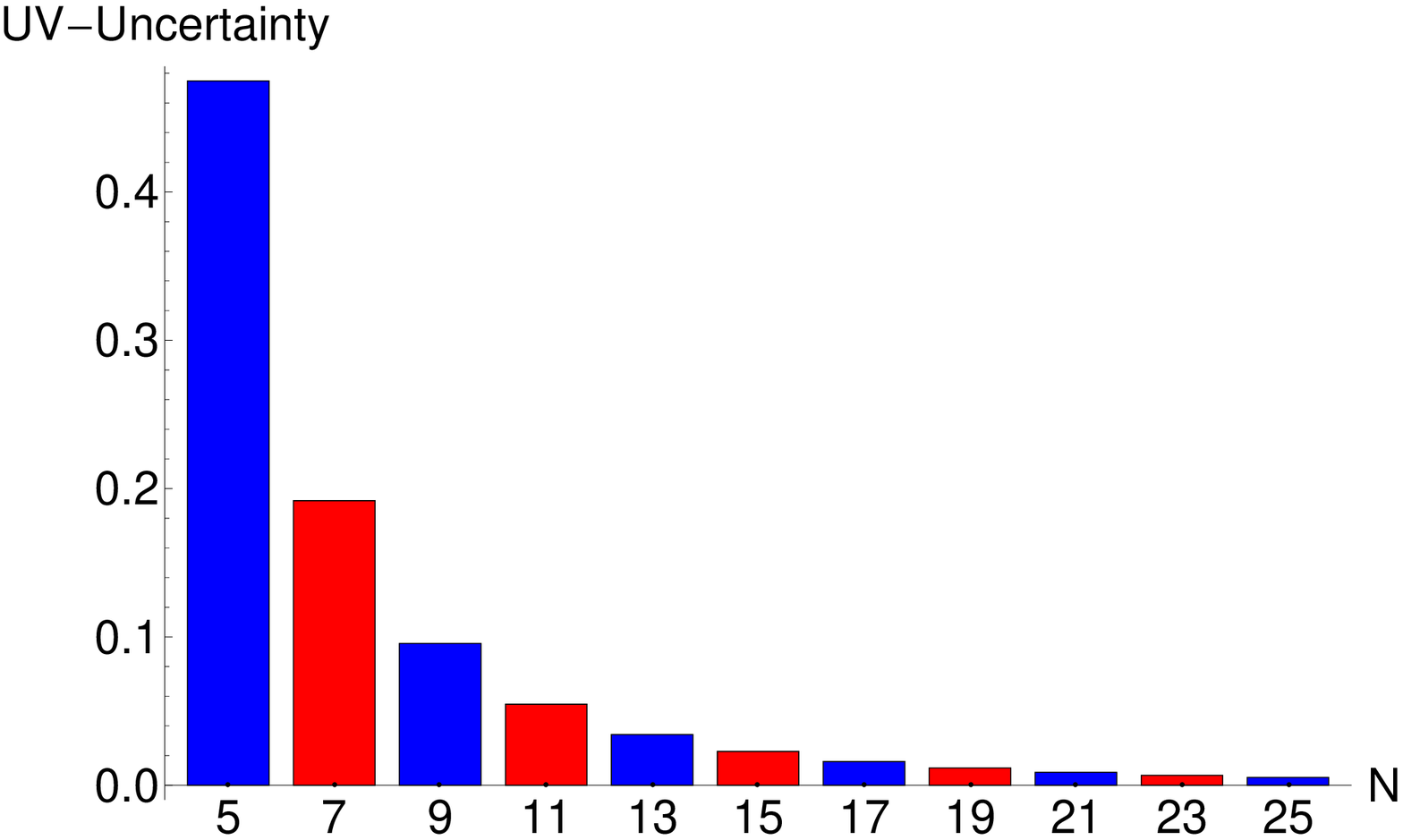}
\end{minipage}
\caption{Plot of $\mathscr{U}_{\innu \innv}$ versus $N \in [5,25]$ for the vacuum state $\lg \mathfrak{q}_{\gamma} | 0 \rg$ defined in
equation (\ref{e15}). Note that $\mathscr{V}_{\innu}$ and $\mathscr{V}_{\innv}$ do not depend on the labels $\mathfrak{p}_{\kappa}$ and
$\mathfrak{q}_{\tau}$ which characterize the $N$-dimensional discrete phase spaces. The $N=3$ case implies in $\mathscr{U}_{\innu \innv}
\approx 1.45$ (exceeds the expected value $\half$) and for this particular reason, such a situation was supressed in the present picture; 
moreover, the asymptotic limit $\mathscr{V}_{\innu} \mathscr{V}_{\innv} \geq \lpar \case{\pi}{N} \rpar^{2}$ can be promptly verified when 
$N \gg 1$.} 
\end{figure}
Figure 2 exhibits the plot of $\mathscr{U}_{\innu \innv}$ versus $N \in [5,25]$ for certain $N$ odd cases, where the normalized density 
operator $\ro$ is here described by the discrete vacuum state (\ref{e15}). Note that the numerical calculations corroborate not only the
inequality (\ref{e29}), but also its respective asymptotic behaviour when $N \gg 1$. However, let us comment some few words on the 
mathematical proof of this theorem given in the supplementary material of reference \cite{Massar}: since this proof basically depends on some
initial insights (which deserve be deeply investigated and tested for different finite quantum states) related to the mean value $\lg \lbk 
{\bf S}_{\innu},{\bf S}_{\innv} \rbk \rg$ --- and its inherent connection with $\lg \lbr {\bf C}_{\innu},{\bf C}_{\innv} \rbr \rg$ --- 
present in equation (\ref{b5}), it is natural to ask whether there exist analogous structures underlying the previous uncertainty principles
showed in appendix B. As this would involve further efforts and detailed studies on important mathematical questions associated with, let
us now spend our time investigating the possible implications of this particular algebraic framework in certain approaches to quantum gravity
with special emphasis on GUP.

\section{The link with quantum gravity}

Various theoretical approaches associated with quantum gravity (\eg, String Theory and Doubly Special Relativity), as well as black hole 
physics suggest, in particular, the existence of a minimum measurable length (or a maximum observable momentum) and/or the breakdown of the 
spacetime continuum, as a natural consequence of combining Quantum Mechanics and General Relativity, whose first implications lead necessarily 
to the so-called GUP. In this prolific scenario, new theoretical frameworks with convenient inherent mathematical properties --- which allow us 
to incorporate some possible gravity effects --- always represent into the literature an important stepforward beyond the mainstream. Thus, 
motivated by a potential application to quantum gravity, we initiate our considerations remembering that ${\it D}_{\mathfrak{q}} = 
\eps^{2-\delta} q_{0}$ and ${\it D}_{\mathfrak{p}} = \eps^{\delta} p_{0}$ correspond to the distances between successive discrete eigenvalues 
of $\bi{Q}$ and $\bi{P}$, with $\delta \in (0,2)$, $q_{0} p_{0} = \hbar$, and $\eps^{2} = \case{2 \pi}{N}$. The connection with the Planck 
length $L_{\inp}$ is reached in this case through the relations ${\it D}_{\mathfrak{q}} = \eps L_{\inp}$ and ${\it D}_{\mathfrak{p}} =
\case{\eps \hbar}{L_{\inp}}$, which implies in distinct forms for $q_{0}$ and $p_{0}$, that is\footnote{Note that $L_{\inp}$ can also be
expressed in terms of well-known physical constants as follows: $L_{\inp}^{2} = \case{G \hbar}{c^{3}}$ ($G$ denotes the gravitational constant, 
while $c$ is the speed of light in the vacuum) and $L_{\inp}^{2} = \lambda_{\com}^{2} \alpha_{\gra}$ ($\lambda_{\com}$ refers to the reduced 
Compton wavelength, $\alpha_{\gra}$ being the dimensionless gravitational coupling constant in such a case). Since $L_{\inp} \approx 1,616
\times 10^{-35} \mathrm{m}$ and $\hbar \approx 1,055 \times 10^{-34} \mathrm{m}^{2} \cdot \mathrm{kg} / \mathrm{seg}$, the ratio 
$\case{\hbar}{L_{\inp}}$ appeared in the different situations of $p_{0}$ is approximately given by $6,528 \, \mathrm{m} \cdot \mathrm{kg} /
\mathrm{seg}$. The Planck mass $M_{\inp} = \sqrt{\case{\hbar c}{G}} \approx 2,176 \times 10^{-8} \mathrm{kg}$ can also be inserted into this
context through the additional link $q_{0} p_{0} = M_{\inp} L_{\inp} c$, which implies in the relation $\case{\hbar}{L_{\inp}} = M_{\inp} c$.},
\brr
\mbox{(i)} \quad \delta = 0 : \quad q_{0} = \frac{L_{\inp}}{\eps} \quad \mbox{and} \quad p_{0} = \frac{\eps \hbar}{L_{\inp}} , \nn \\
\mbox{(ii)} \quad \delta = 1 : \quad q_{0} = L_{\inp} \quad \mbox{and} \quad p_{0} = \frac{\hbar}{L_{\inp}} , \nn \\
\mbox{(iii)} \quad \delta = 2 : \quad q_{0} = \eps L_{\inp} \quad \mbox{and} \quad p_{0} = \frac{\hbar}{\eps L_{\inp}} . \nn
\err
Let us now analyze, in a first moment, the effect of these connections on $\mathscr{U}_{\innu \innv}$ through proper expansions of the unitary 
operators defined by means of equation (\ref{e20}).

For instance, the power-series expansions related to the mean values $\lg {\bf U} \rg$ and $\lg {\bf V} \rg$, namely,
\bd 
\fl \quad \lg {\bf U} \rg = \sum_{k=0}^{\infty} \frac{(-1)^{k}}{(2k)!} \lpar \frac{\eps}{L_{\inp}} \rpar^{2k} \lg \bi{Q}^{2k} \rg + \rmi
\sum_{k=0}^{\infty} \frac{(-1)^{k}}{(2k+1)!} \lpar \frac{\eps}{L_{\inp}} \rpar^{2k+1} \lg \bi{Q}^{2k+1} \rg
\ed
and
\bd
\fl \quad \lg {\bf V} \rg = \sum_{k=0}^{\infty} \frac{(-1)^{k}}{(2k)!} \lpar \frac{\eps L_{\inp}}{\hbar} \rpar^{2k} \lg \bi{P}^{2k} \rg + \rmi
\sum_{k=0}^{\infty} \frac{(-1)^{k}}{(2k+1)!} \lpar \frac{\eps L_{\inp}}{\hbar} \rpar^{2k+1} \lg \bi{P}^{2k+1} \rg ,
\ed
lead us to obtain approximate expressions for $\mathscr{V}_{\innu(\innv)}$ as follow:
\brr
\lb{e30}
\fl \quad \mathscr{V}_{\innu} \approx \lpar \frac{\eps}{L_{\inp}} \rpar^{2} \mathscr{V}_{\inq} - \lpar \frac{\eps}{L_{\inp}} \rpar^{4} 
\lpar \frac{\lg \bi{Q}^{4} \rg}{12} + \frac{\lg \bi{Q}^{2} \rg^{2}}{4} - \frac{\lg \bi{Q} \rg \lg \bi{Q}^{3} \rg}{3} \rpar + \cdots \; , \\
\lb{e31}
\fl \quad \mathscr{V}_{\innv} \approx \lpar \frac{\eps L_{\inp}}{\hbar} \rpar^{2} \mathscr{V}_{\inp} - \lpar \frac{\eps L_{\inp}}{\hbar} \rpar^{4} 
\lpar \frac{\lg \bi{P}^{4} \rg}{12} + \frac{\lg \bi{P}^{2} \rg^{2}}{4} - \frac{\lg \bi{P} \rg \lg \bi{P}^{3} \rg}{3} \rpar + \cdots \; . 
\err
Note that both expressions are written in terms of even powers of the coefficients $\case{\eps}{L_{\inp}}$ and $\case{\eps L_{\inp}}{\hbar}$, the
leading terms being directly associated with the respective variances $\mathscr{V}_{\inq}$ and $\mathscr{V}_{\inp}$; in addition, the first 
subleading terms are interpreted in this context as corrections whose effectiveness depends on the quantum state $\ro$ under investigation. Next, 
let us mention briefly that the expected asymptotic behaviour of $\mathscr{U}_{\innu \innv}$ for $N \gg 1$ can be properly estimated from relation
\bd
\frac{\mathscr{V}_{\innu} \mathscr{V}_{\innv}}{A^{2}} \geq 1 - \lpar \mathscr{V}_{\innu} + \mathscr{V}_{\innv} + \frac{2}{A} \mathscr{V}_{\innu}
\mathscr{V}_{\innv} \rpar ,
\ed
since the terms in parenthesis on the right-hand side are negligible in front of $1$ --- in such a case, $\mathscr{V}_{\innu}$ and $\mathscr{V}_{\innv}$
are both sufficiently small. So, after some minor adjustments in our calculations, it is possible to derive the generalized uncertainty principle
\brr
\lb{e32}
\fl \mathscr{V}_{\inq} \mathscr{V}_{\inp} \geq & \frac{\hbar^{2}}{4} \left\{ 1 - \left[ \lpar \frac{\eps}{L_{\inp}} \rpar^{2} \mathscr{V}_{\inq} - 
\lpar \frac{\eps}{L_{\inp}} \rpar^{4} \lpar \frac{\lg \bi{Q}^{4} \rg}{12} + \frac{\lg \bi{Q}^{2} \rg^{2}}{4} - \frac{\lg \bi{Q} \rg \lg \bi{Q}^{3} \rg}{3} 
\rpar \right. \right. \nn \\
\fl & + \left. \left. \lpar \frac{\eps L_{\inp}}{\hbar} \rpar^{2} \mathscr{V}_{\inp} - \lpar \frac{\eps L_{\inp}}{\hbar} \rpar^{4} \lpar 
\frac{\lg \bi{P}^{4} \rg}{12} + \frac{\lg \bi{P}^{2} \rg^{2}}{4} - \frac{\lg \bi{P} \rg \lg \bi{P}^{3} \rg}{3} \rpar + \cdots \right] \right\} , 
\err
which exhibits additional terms if one compares with the usual Heisenberg-Kennard-Robertson inequality \cite{HKR}. Besides, once the expansions 
involved in this evaluation are locally valid for a predetermined region of the $N^{2}$-dimensional discrete phase space, it turns natural to ask
whether periodic boundary conditions are present or not in $\ro$ with the simple aim of avoiding multivalued mean values of non-periodic operators
like $\bi{Q}$ and $\bi{P}$.

The applications of this particular algebraic approach can also be extended to the results established in the appendix B. Indeed, the four RS
uncertainty principles inherent to the sine and cosine operators --- which are defined in terms of the unitary operators ${\bf U}$ and ${\bf V}$ ---
certainly represent an ideal scenario where new corrections are obtained, and consequently, new generalized uncertainty principles are derived. Along
this specific research line, some additional theoretical and numerical investigations involving certain realistic physical systems (prepared in
different initial quantum states) would also be highly desirable into this context.   

In summary, the theoretical framework developed until now has provided a new approach to see explicitly corrections (in the form of a GUP) that are
naturally present in a genuine quantum theory emerged from a finite discrete configuration space. With this in mind, let us make some comments
on the recent work of Bang and Berger \cite{Berger}, where a similar theoretical framework has been employed: the particular uncertainty relation
$\mathscr{V}_{\innu} \mathscr{V}_{\innv} \geq | \lg \delta {\bf V}^{\dagger} \delta {\bf U} \rg |^{2}$ represents an important starting point in
their formalism and whose veracity was not properly checked (or even reported) in subsequent papers. Despite the apparent simplicity of this problem, 
the work, per se, has its merits and the GUP derived from this relation shows the `expected corrections' written in terms of the Plack length
$L_{\inp}$. In many ways the analysis here presented of the generalized uncertainty principle is complementary to that provided in Ref. \cite{Berger}; 
besides, in what concerns the Planck scale and its implications for different theoretical approaches, our results sound promising at first glance. 

\section{Concluding remarks}

In this paper, we have constructed, from first principles, a self-consistent formalism for treating physical systems described by a finite space 
of states which exhibits, within other features, potential applications in quantum gravity. For this particular task, we initially introduce the
displacement generator $\{ {\bf D}(\eta,\xi) \}_{\eta,\xi = -\ell, \ldots, \ell}$ in a $N^{2}$-dimensional discrete phase space labelled by the 
dual momentumlike and coordinatelike variables $\eta$ and $\xi$ which obey the arithmetic modulo $N$, for then establishing a set of relevant 
intrinsic properties where the parity operator $\opp$ represents an important element of this theoretical approach. Indeed, the unitary transformation 
${\bf D}(\mu,\nu) \opp {\bf D}^{\dagger}(\mu,\nu)$ on the parity operator has allowed us to define the mod($N$)-invariant unitary operator basis 
$\{ \bDelta(\mu,\nu) \}_{\mu,\nu = -\ell, \ldots,\ell}$, whose rich algebraic structure was extensively explored and discussed along the text. In 
this sense, it is worth stressing that the connection between $\bDelta(\mu,\nu)$ and ${\bf D}(\eta,\xi)$ through a discrete Fourier transform can be
interpreted as a fundamental feature of this construction process that reinforces such a structure.

In what concerns the complete orthonormal operator basis $\bDelta(\mu,\nu)$, any linear operator ${\bf O}$ can now be promptly decomposed in this
basis and its coefficients present a one-to-one correspondence between operators and functions belonging to a well-defined $N^{2}$-dimensional 
discrete phase space. The first immediate consequence of this result refers to the mean value $\lg {\bf O} \rg = \Tr [ {\bf O} \ro ]$, since
it can be expressed in terms of a simple mathematical operation involving the product of the discrete Wigner function $\mathscr{W}_{\rho}(\mu,\nu)$
--- interpreted in this context as a weight function --- and the coefficients $\mathscr{O}(\mu,\nu)$. The second one yields the mean values of
the commutation and anticommutation relations of two non-commuting operators, whose respective mapped expressions preserve the embryonic structure
of the sine and cosine functions present in the continuous phase-space counterpart \cite{Leaf}. In fact, this important result reveals the
presympletic structure of geometric origin inherent to a finite-dimensional discrete phase space with toroidal topology. The third natural
consequence of this constructive process, which is directly associated with the discrete mapping kernel $\bDelta(\mu,\nu)$, allows us to study
the dynamics of any isolated physical system $\ro(t)$ described by a time-independent Hamiltonian and whose space of states is finite. So, the 
mapped equivalent of the von Neumann-Liouville equation leads to obtain a differential equation for the discrete Wigner function where, in principle,
the formal solution for $\mathscr{W}_{\rho}(\mu,\nu;t)$ can be promptly established. As usual, this solution basically shows the time evolution of a 
given initial Wigner function by means of a time-dependent discrete phase-space propagator here expressed as interated applications of the mapped
Liouville operator upon $\mathscr{W}_{\rho}(\alf,\bet;t_{0})$. Following, as a first illustration of this formal solution, we have investigated the
discrete analogue of the harmonic oscillator, via Harper Hamiltonian, with the aim of obtaining a differential equation for the time-dependent
propagator $\mathscr{P}_{\inh}(\mu,\nu;t | \alf,\bet;t_{0})$ --- see equation (\ref{e14}). Within our theoretical framework, this last result
still represents an open window of future numerical investigations since there exist different models of finite harmonic oscillator in literature
with distinct mathematical features \cite{MOHF}. 

The discrete coherent-state projector (\ref{e16}) certainly represents, into the algebraic approach here developed, an important example of finite 
quantum states with periodic boundary conditions since it embodies certain mathematical properties inherent to the Jacobi theta functions \cite{WW}. 
Furthermore, two fundamental properties related to the discrete coherent states (namely, the non-orthogonality and completeness relations) were also
discussed in the body of the text. It is worth mentioning that the underlying simplicity of ${\bf D}(\kappa,\tau)$ is in sharp contrast with that
definition employed in \cite{Galetti1} for the discrete displacement generator --- in this case, $\dis(\kappa,\tau) \coloneq \om^{-\half \kappa \tau}
{\bf U}^{\kappa} {\bf V}^{-\tau}$ --- once it intrinsically possesses modulo $N$ symmetry with important theoretical implications (even regarding
numerical calculations). In this sense, future studies involving detailed analyses on the possible mathematical effects of each particular description
in different finite physical systems will be welcome, and will certainly represent a stepforward in the crucial construction process of a solid
theoretical framework for such systems with an underlying discrete space.   

Now, let us discuss the different scenarios associated with the uncertainty principle for finite-dimensional discrete phase spaces. The first natural
scenario corresponds to the well-established discrete coordinate and momentum Hermitian operators and their inherent RS uncertainty principle. For this
purpose, we have introduced the operators $\bi{Q}$ and $\bi{P}$ by means of eigenvalue equations which present certain peculiarities: the respective
eigenvalues $\mathfrak{q}_{\iaf} \coloneq {\it D}_{\mathfrak{q}} \alf$ and $\mathfrak{p}_{\ibe} \coloneq {\it D}_{\mathfrak{p}} \bet$ are here defined
as multiplicative factors of the basic quantities ${\it D}_{\mathfrak{q}} = \eps^{2-\delta} q_{0}$ and  ${\it D}_{\mathfrak{p}} = \eps^{\delta} p_{0}$
--- which are responsible by the distances between sucessive eigenvalues of $\bi{Q}$ and $\bi{P}$, that is, $\mathfrak{q}_{\alf + 1} -
\mathfrak{q}_{\alf} = {\it D}_{\mathfrak{q}}$ and $\mathfrak{p}_{\bet+1} - \mathfrak{p}_{\bet} = {\it D}_{\mathfrak{p}}$ for all $\{ \alf,\bet \} \in
[-\ell,\ell]$ --- with $\delta \in (0,2)$ and $q_{0} p_{0} = \hbar$ fixed. Since $\eps = \sqrt{\case{2 \pi}{N}}$, it turns immediate to obtain from
this context the standard spectra related to the continuous counterparts of these operators in the limit $N \rightarrow \infty$. This particular
definition then allows us to establish a first realization for the unitary operators with immediate theoretical implications. Indeed, expressions for
mean values of moments as well as mean values of commutation and anticommutation relations related to the discrete coordinate and momentum operators
can now be properly evaluated for any finite physical system where periodic boundary conditions does not apply. Thus, the first part culminates with
the discussion on the RS uncertainty principle and its inherent limitations.

The second scenario basically works out with unitary operators defined as complex exponentials whose arguments are written in terms of $\bi{Q}$ 
and $\bi{P}$ --- see equation (\ref{e20}). Such a mathematical procedure has the particular virtues of: (i) avoiding multivalued mean values,
(ii) searching for new uncertainty relations associated with such operators, and finally (iii) describing a wide class of physical systems with finite
space of states where the periodic boundary conditions are present. The constraint between correlation function $\lg {\bf V}^{\xi} \rg$ and discrete
Fourier transform of the intensity time series $\lg {\bf U}^{\eta} \rg$, here expressed through a theorem due to Massar and Spindel \cite{Massar},
certainly represents the highest point of this constructive process since it leads us to establish a bound for the variances $\mathscr{V}_{\innu}$
and $\mathscr{V}_{\innv}$. The next stage consists in discussing under what circumstances the Planck scale $L{\inp}$ --- or even the Planck mass
$M_{\inp}$ --- can be inserted into this algebraic approach.

A possible connection with $L_{\inp}$ and $M_{\inp}$ is here reached by fixing the basic distances ${\it D}_{\mathfrak{q}}$ and 
${\it D}_{\mathfrak{p}}$ as follow: ${\it D}_{\mathfrak{q}} = \eps L_{\inp}$ and ${\it D}_{\mathfrak{p}} = \eps M_{\inp} c$ $\lpar \mbox{or} \; 
{\it D}_{\mathfrak{p}} = \case{\eps \hbar}{L_{\inp}} \rpar$. Thus, the expansion of the mean values $\lg {\bf U} \rg$ and $\lg {\bf V} \rg$ written in
terms of even and odd powers associated with the respective ratios $\case{\eps}{L_{\inp}}$ and $\case{\eps L_{\inp}}{\hbar}$ allows us, in particular,
to evaluate approximate expressions for $\mathscr{V}_{\innu}$ and $\mathscr{V}_{\innv}$. It is important to stress that both the expressions already
embody well-known terms and corrections, whose effectiveness depends essentially on the quantum state $\ro$. The next natural step consists in
employing such approximations into equation (\ref{e29}) in order to obtain a uncertainty principle with unique features: the first-order term resembles
the usual Heisenberg-Kennard-Robertson inequality \cite{HKR}, namely $\mathscr{V}_{\inq} \mathscr{V}_{\inp} \geq \case{\hbar^{2}}{4}$, while the
additional terms represent important corrections that improve the previous contribution. This particular GUP, nevertheless, presents certain
mathematical restrictions since its validity depends upon a predetermined region of the finite-dimensional discrete phase space where the power-series
expansions can be achieved. With this in mind, let us now briefly mention about an alternative link with the previous Planck units $L_{\inp}$ and
$M_{\inp}$ aiming the experimental fit. For instance, if one considers $q_{0} = \mathfrak{s} L_{\inp}$ and $p_{0} = \case{\hbar}{\mathfrak{s} L_{\inp}}$
(or $p_{0} = \mathfrak{s}^{-1} M_{\inp} c$) fixed with $\mathfrak{s} \in \mathbb{R}_{+}^{\ast}$, the distances ${\it D}_{\mathfrak{q}}$ and 
${\it D}_{\mathfrak{p}}$ assume, respectively, the different forms $\mathfrak{s} \eps^{2-\delta} L_{\inp}$ and $\case{\eps^{\delta} \hbar}
{\mathfrak{s} L_{\inp}}$, which preserve, in such a case, the relations $q_{0} p_{0} = \hbar$ and $\case{{\it D}_{\mathfrak{p}} {\it D}_{\mathfrak{q}}}
{2 \pi \hbar} = \case{1}{N}$. Note that $\delta$ and $\mathfrak{s}$ represent two important `free parameters' which can be used to validade equation
(\ref{e32}) by means of experimental data. From the theoretical point of view, the parameter $\mathfrak{s}$ is responsible, in principle, for squeezing
effects in finite-dimensional spaces \cite{Marchiolli1}.

In a more pragmatic sense, it is worth stressing that the compilation of results here presented not only corroborates and generalizes those obtained 
in current literature, but also represents a concatenated effort in joining two promising research branches of physics devoted to the study of quantum 
information theory and quantum gravity. Note that the huge difficulties in solving the intriguing problem related to breakdown of the spacetime
continuum still remain the same, since the focus of this paper consists in exhibiting a self-consistent theoretical framework for the quantum-gravity
community where certain algebraic structures inherent to finite-dimensional discrete phase spaces and underlying generalized uncertainty principles can
be worked out without apparent problems. Besides, our results also seem to be quite suitable to deal with a wide range of problems related to quantum
computing, statistical mechanics, and also foundations of quantum mechanics. Finally, let us mention that there exists another path for future research
which will be properly presented in due course.

\ack

In this occasion, it is both a privilege and a pleasure to dedicate this specific paper to Professor Di\'{o}genes Galetti, senior theoretical 
physicist from the Instituto de F\'{\i}sica Te\'{o}rica (Universidade Estadual Paulista, SP, Brazil), colleague and friend. Moreover, it is 
also an opportune moment to appreciate the insights and fruitful discussions we gained during all those productive years from his pioneering 
contributions in Quantum Mechanics and Nuclear Physics. The authors thank the local organizing committee (in particular, the chairman Salomon 
Sylvain Mizrahi) for partial financial support in the `{\sl $12^{{\rm th}}$ International Conference on Squeezed States and Uncertainty 
Relations}' (2-6 May 2011, Foz do Igua\c{c}u, Brazil), where this work was initially presented. We also appreciate the helpful discussions of 
our colleague Daniel Jonathan (Universidade Federal Fluminense, RJ, Brazil) on highly pertinent questions related to this work.

\appendix
\section{On the Wigner function for the discrete coherent states}

Let us initiate this mathematical appendix remembering that equation (\ref{e15}) represents a particular case where the wavefunction in the
position-like representation for the discrete coherent states is involved, namely,
\be
\lb{a1}
\fl \qquad \qquad \lg u_{\gam} | \eta,\xi \rg = \lbk \frac{2 \mathfrak{a}}{\mathscr{M}(0,0)} \rbk^{\half} \om^{- \{ 2^{-1} \eta \xi \} + 
\gam \eta} \; \vartheta_{3} \lpar 2 \mathfrak{a} (\gam - \xi) | 2 \rmi \mathfrak{a} \rpar .
\ee
This particular wavefunction has a central role into our main purpose since the mean value $\lg \eta,\xi | \opp | \eta,\xi \rg$ is connected
directly with the expression
\bd
\fl \qquad \lg \eta,\xi | \opp | \eta,\xi \rg = \frac{2 \mathfrak{a}}{\mathscr{M}(0,0)} \sum_{\gam = - \ell}^{\ell} \om^{2 \gam \eta} \,
\vartheta_{3} \lpar 2 \mathfrak{a} (\gam - \xi) | 2 \rmi \mathfrak{a} \rpar \vartheta_{3} \lpar 2 \mathfrak{a} (\gam + \xi) | 2 \rmi 
\mathfrak{a} \rpar .
\ed
The next step consists in replacing the product of Jacobi $\vartheta_{3}$-functions by the respective product of two infinite sums in order
to perform the sum over the discrete variable $\gam$,
\be
\lb{a2}
\fl \qquad \qquad \lg \eta,\xi | \opp | \eta,\xi \rg = \mathscr{M}^{-1}(0,0) \exp \lbk - \frac{4 \pi}{N} \eta (\eta + \rmi \xi) \rbk 
\mathscr{F}(\eta,\xi)
\ee
where
\bd
\fl \quad \mathscr{F}(\eta,\xi) = \sum_{ \{ \alf,\bet \} \in \mathbb{Z}} \exp \lbk - \frac{2 \pi}{N} \lpar \alf - \frac{N \bet}{2} 
\rpar^{2} - \frac{\pi N}{2} \bet^{2} + 4 \pi \eta \bet - \frac{4 \pi}{N} \alf (\eta + \rmi \xi) \rbk .
\ed
Consequently, the sum over $\bet$ can now be separated out in two contributions coming from the even $(e)$ and odd $(o)$ integers:
$\mathscr{F} = \mathscr{F}_{e} + \mathscr{F}_{o}$. In the following, let us go one step further with the main aim of determining each term 
separately. 

For instance, the even term $\mathscr{F}_{e}$ can be dealt with by shifting the sum over $\alf$ by $N \bet$. This specific trick then
produces a decoupling between the discrete labels $\alf$ and $\bet$, that is
\bd
\fl \quad \mathscr{F}_{e}(\eta,\xi) = \sum_{\alf \in \mathbb{Z}} \exp \lbk - \frac{2 \pi}{N} \alf^{2} - \frac{4 \pi}{N} \alf (\eta + \rmi \xi) 
\rbk \sum_{\bet \in \mathbb{Z}} \exp \lbk - 2 \pi N \bet^{2} + 4 \pi \bet (\eta - \rmi \xi) \rbk ,
\ed
which permits us to identify each sum with a particular Jacobi $\vartheta_{3}$-function \cite{WW}, 
\be
\lb{a3}
\fl \qquad \qquad \mathscr{F}_{e}(\eta,\xi) = \sqrt{\mathfrak{a}} \, \exp \lbk \frac{4 \pi}{N} \eta (\eta + \rmi \xi) \rbk \vartheta_{3} 
( 2 \mathfrak{a} \eta | \rmi \mathfrak{a} ) \vartheta_{3} ( 4 \mathfrak{a} \xi | 4 \rmi \mathfrak{a} )
\ee
with $\mathfrak{a} \equiv (2N)^{-1}$. The odd term can also be evaluated through a similar mathematical procedure, yielding as result the 
closed-form expression
\be
\lb{a4}
\fl \qquad \qquad \mathscr{F}_{o}(\eta,\xi) = \sqrt{\mathfrak{a}} \, \exp \lbk \frac{4 \pi}{N} \eta (\eta + \rmi \xi) \rbk \vartheta_{4} 
( 2 \mathfrak{a} \eta | \rmi \mathfrak{a} ) \vartheta_{2} ( 4 \mathfrak{a} \xi | 4 \rmi \mathfrak{a} ) ,
\ee
where now the $\vartheta_{4}$-  and $\vartheta_{2}$-functions are involved. Thus, substituting these terms into equation (\ref{a2}), 
we finally obtain $\lg \eta,\xi | \opp | \eta,\xi \rg = \mathscr{M}(2 \eta, 2 \xi)/ \mathscr{M}(0,0)$. In this case, the real auxiliary
function 
\bd
\mathscr{M}(2 \eta, 2 \xi) = \exp \lbk - \frac{4 \pi}{N} \eta (\eta + \rmi \xi) \rbk \mathscr{F}(\eta,\xi) 
\ed
denotes the sum of products of Jacobi theta functions evaluated at integer arguments. As an immediate result, the discrete Wigner function
(\ref{e17}) can be promptly determined when $\eta = \kappa - \mu$ and $\xi = \tau - \nu$.

Next, let us derive the marginal distributions associated with the discrete Wigner function $\mathscr{W}_{\kappa,\tau}(\mu,\nu)$ through 
the standard mathematical prodecure
\bd
\fl \quad \mathscr{Q}_{\tau}(\nu) \equiv \frac{1}{N} \sum_{\mu = - \ell}^{\ell} \mathscr{W}_{\kappa,\tau}(\mu,\nu) \qquad
\mbox{and} \qquad \mathscr{R}_{\kappa}(\mu) \equiv \frac{1}{N} \sum_{\nu = - \ell}^{\ell} \mathscr{W}_{\kappa,\tau}(\mu,\nu) .
\ed
Thus, after some calculations based on the results obtained in \cite{WW} for the Jacobi theta functions, it is easy to reach the
closed-form expressions
\brr
\lb{a5}
\fl \; \mathscr{Q}_{\tau}(\nu) = \sqrt{4 \mathfrak{a}} \, \frac{\vartheta_{3}(0|4 \rmi \mathfrak{a}) \vartheta_{3}(4 \mathfrak{a} (\tau - \nu)
| 4 \rmi \mathfrak{a}) + \vartheta_{2}(0|4 \rmi \mathfrak{a}) \vartheta_{2}(4 \mathfrak{a} (\tau - \nu)| 4 \rmi \mathfrak{a})}{\vartheta_{3}
(0| \rmi \mathfrak{a}) \vartheta_{3}(0|4 \rmi \mathfrak{a}) + \vartheta_{4}(0| \rmi \mathfrak{a}) \vartheta_{2}(0|4 \rmi \mathfrak{a})}
\err
and
\brr
\lb{a6}
\fl \; \mathscr{R}_{\kappa}(\mu) = \sqrt{\mathfrak{a}} \, \frac{\vartheta_{3}(0| \rmi \mathfrak{a}) \vartheta_{3}(2 \mathfrak{a} (\kappa - \mu)
| \rmi \mathfrak{a}) + \vartheta_{4}(0| \rmi \mathfrak{a}) \vartheta_{4}(2 \mathfrak{a} (\kappa - \mu)| \rmi \mathfrak{a})}{\vartheta_{3}
(0| \rmi \mathfrak{a}) \vartheta_{3}(0|4 \rmi \mathfrak{a}) + \vartheta_{4}(0| \rmi \mathfrak{a}) \vartheta_{2}(0|4 \rmi \mathfrak{a})} .
\err
It is worth emphasizing that $\mathscr{W}_{\kappa,\tau}(\mu,\nu)$ and its respective marginal distributions are normalized to unity in such a 
case, namely,
\bd
\frac{1}{N} \sum_{\mu,\nu = -\ell}^{\ell} \mathscr{W}_{\kappa,\tau}(\mu,\nu) = \sum_{\nu = -\ell}^{\ell} \mathscr{Q}_{\tau}(\nu) =
\sum_{\mu = - \ell}^{\ell} \mathscr{R}_{\kappa}(\mu) = 1 . 
\ed
\begin{figure}[!t]
\centering
\begin{minipage}[b]{0.45\linewidth}
\includegraphics[width=\textwidth]{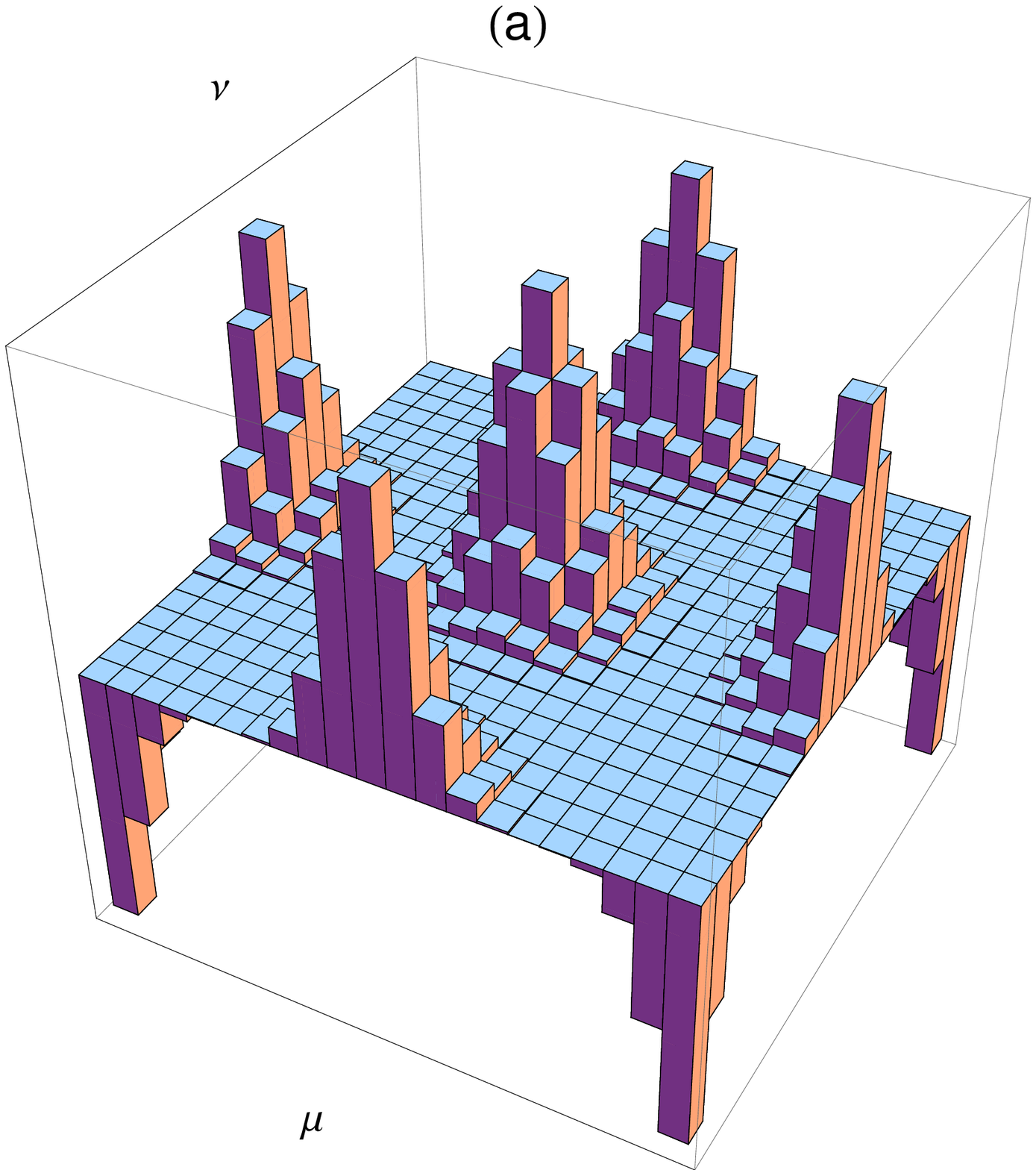}
\end{minipage} \hfill
\begin{minipage}[b]{0.45\linewidth}
\includegraphics[width=\textwidth]{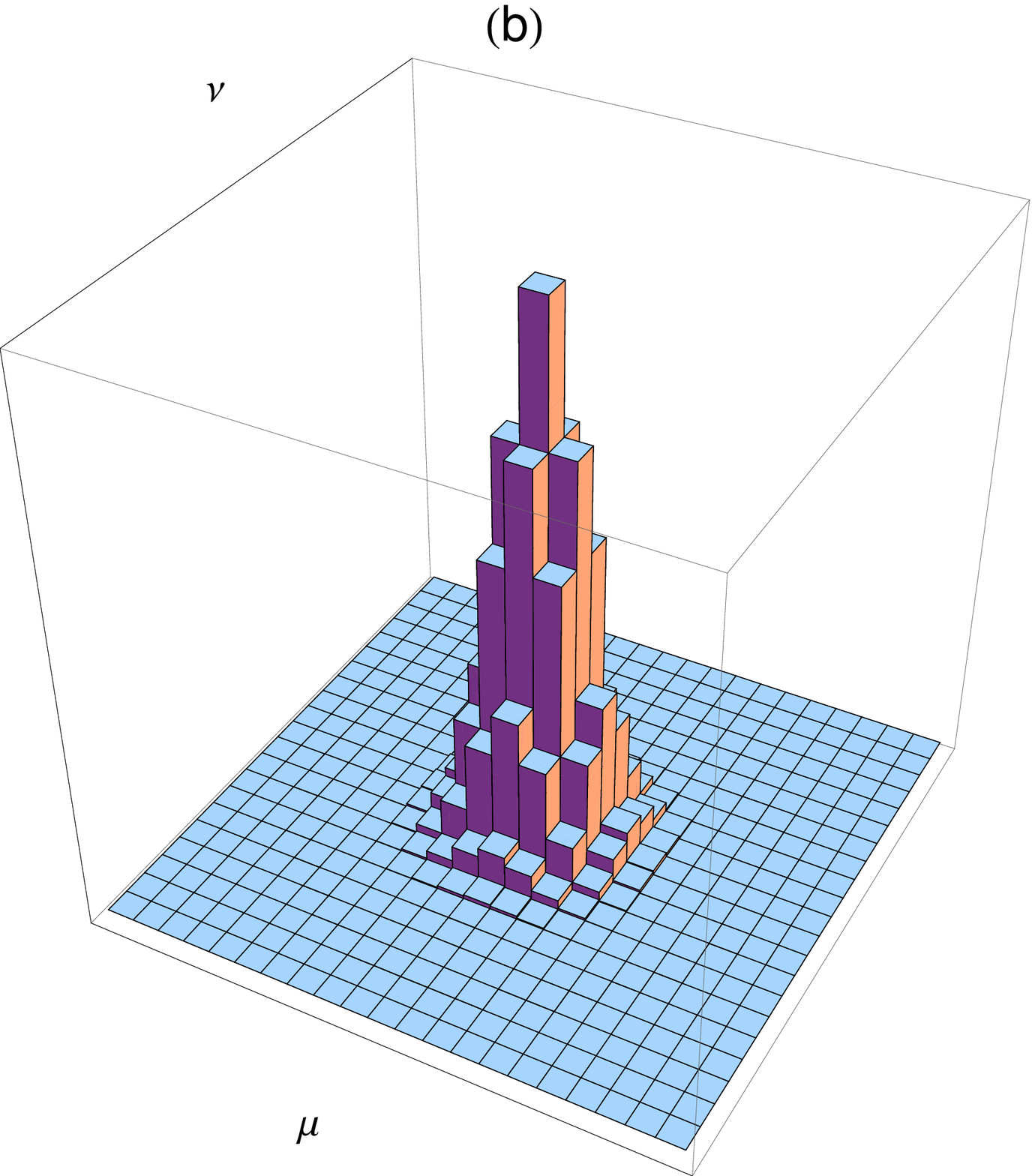}
\end{minipage}
\caption{Three-dimensional plots of (a) $\mathscr{W}_{0,0}(\mu,\nu)$ and (b) $\mathscr{R}_{0}(\mu) \mathscr{Q}_{0}(\nu)$ versus 
$\{ \mu,\nu \} \in [-10,10]$ for the quantum vacuum state, with $N=21$ fixed. Note that in (b) both the marginal distributions assume 
strictly positive values, exhibiting, as expected, a gaussian profile localized in the central part of the $21^{2}$-dimensional phase 
space labeled by the dimensionless discrete variables $\mu$ and $\nu$. In this way, the partial sums --- verified in the definition of 
the marginal distributions --- over the discrete variables $\mu$ and $\nu$ `wash' the negative values of $\mathscr{W}_{0,0}(\mu,\nu)$ 
showed in (a), namely, the quantum correlations inherent to the vacuum state disappear in such a mathematical construction.} 
\end{figure}
%
Now, let us adopt the vacuum state $| \kappa , \tau \rg \equiv | 0,0 \rg$ as a reference state in our numerical calculations. Figure A1 
exhibits the 3D plots of (a) $\mathscr{W}_{0,0}(\mu,\nu)$ and (b) $\mathscr{R}_{0}(\mu) \mathscr{Q}_{0}(\nu)$ as a function of $\{ \mu,\nu \}
\in [-10,10]$ with $N=21$ fixed. The negative values appeared in (a) represent a quantum signature of the nonclassical effects related to 
the reference state under investigation. Besides, it is worth mentioning that such a behaviour disappears in (b) and this fact is directly 
associated with the nonexistence of correlations between the discrete variables $\mu$ and $\nu$, \ie, $\mathscr{W}_{0,0}(\mu,\nu) \neq
\mathscr{R}_{0}(\mu) \mathscr{Q}_{0}(\nu)$ \cite{Ferrie}.

\section{The RS uncertainty principles for sine and cosine operators}

We initiate this appendix by defining four elements of an important set of Hermitian operators (recognized in the literature by sine and 
cosine operators), whose intrinsic properties are promptly explored within a specific theoretical point of view where the RS uncertainty 
principle has a central role. Thus, adopting as reference quantum state the discrete coherent states, we then establish some essential 
closed-form results which permit us to analyse the uncertainty principles related to certain relevant combinations of non-commuting elements.
Basically, such results constitute the initial mathematical steps necessary, hitherto, to the development of new inequalities associated 
with the unitary operators ${\bf U}$ and ${\bf V}$.

\subsubsection*{{\bf Definition.}}
{\it Let $\lbr {\bf C}_{\innu},{\bf S}_{\innu},{\bf C}_{\innv},{\bf S}_{\innv} \rbr$ denote four Hermitian operators defined in terms of 
simple combinations of the unitary operators ${\bf U}$ and ${\bf V}$ as follow:}
\be
\lb{b1}
\fl {\bf C}_{\innu} \coloneq \frac{{\bf U} + {\bf U}^{\dagger}}{2} , \quad {\bf S}_{\innu} \coloneq \frac{{\bf U} - {\bf U}^{\dagger}}{2 \rmi} ,
\quad {\bf C}_{\innv} \coloneq \frac{{\bf V} + {\bf V}^{\dagger}}{2}, \quad {\bf S}_{\innv} \coloneq \frac{{\bf V} - {\bf V}^{\dagger}}{2 \rmi} .
\ee
{\it The commutation relations involving these operators present a direct connection with certain anticommutation relations, that is,}
\brr
\fl \qquad \lbk {\bf C}_{\innu},{\bf C}_{\innv} \rbk = \rmi \tan \lpar \case{\pi}{N} \rpar \lbr {\bf S}_{\innu},{\bf S}_{\innv} \rbr , \quad
\lbk {\bf C}_{\innu},{\bf S}_{\innv} \rbk = - \rmi \tan \lpar \case{\pi}{N} \rpar \lbr {\bf S}_{\innu},{\bf C}_{\innv} \rbr , \nn \\
\fl \qquad \lbk {\bf S}_{\innu},{\bf C}_{\innv} \rbk = - \rmi \tan \lpar \case{\pi}{N} \rpar \lbr {\bf C}_{\innu},{\bf S}_{\innv} \rbr , \quad
\lbk {\bf S}_{\innu},{\bf S}_{\innv} \rbk = \rmi \tan \lpar \case{\pi}{N} \rpar \lbr {\bf C}_{\innu},{\bf C}_{\innv} \rbr . \nn
\err
{\it In principle, such results lead us to conclude that partial information on two particular elements of the set is not possible since 
the complementary elements are also necessary into this context; consequently, complete information on the unitary operators is shared in 
four RS uncertainty principles (which will be discussed below) for each different pair of relevant operators. Besides, the additional result
${\bf C}_{\innu}^{2} + {\bf S}_{\innu}^{2} = {\bf C}_{\innv}^{2} + {\bf S}_{\innv}^{2} = {\bf 1}$ resembles a well-established mathematical
property associated with two important trigonometric functions: the sine and cosine functions. Henceforth such Hermitian operators will be 
termed by sine and cosine operators.}

The commutation relations underlying to the sine and cosine operators permit to establish four RS uncertainty principles within this particular
theoretical approach:
\brr
\lb{b2}
\mathscr{U}_{\inc_{\innu} \inc_{\innv}} \coloneq \mathscr{V}_{\inc_{\innu}} \mathscr{V}_{\inc_{\innv}} - \lpar
\mathscr{C}_{\inc_{\innu} \inc_{\innv}} \rpar^{2} \geq \case{1}{4} \left| \lg \lbk {\bf C}_{\innu},{\bf C}_{\innv} \rbk \rg \right|^{2} , \\
\lb{b3}
\mathscr{U}_{\inc_{\innu} \ins_{\innv}} \coloneq \mathscr{V}_{\inc_{\innu}} \mathscr{V}_{\ins_{\innv}} - \lpar
\mathscr{C}_{\inc_{\innu} \ins_{\innv}} \rpar^{2} \geq \case{1}{4} \left| \lg \lbk {\bf C}_{\innu},{\bf S}_{\innv} \rbk \rg \right|^{2} , \\
\lb{b4}
\mathscr{U}_{\ins_{\innu} \inc_{\innv}} \coloneq \mathscr{V}_{\ins_{\innu}} \mathscr{V}_{\inc_{\innv}} - \lpar
\mathscr{C}_{\ins_{\innu} \inc_{\innv}} \rpar^{2} \geq \case{1}{4} \left| \lg \lbk {\bf S}_{\innu},{\bf C}_{\innv} \rbk \rg \right|^{2} , \\
\lb{b5}
\mathscr{U}_{\ins_{\innu} \ins_{\innv}} \coloneq \mathscr{V}_{\ins_{\innu}} \mathscr{V}_{\ins_{\innv}} - \lpar
\mathscr{C}_{\ins_{\innu} \ins_{\innv}} \rpar^{2} \geq \case{1}{4} \left| \lg \lbk {\bf S}_{\innu},{\bf S}_{\innv} \rbk \rg \right|^{2} ,
\err
where the respective covariance functions $\mathscr{C}$'s are given by
\brr
\fl \quad \mathscr{C}_{\inc_{\innu} \inc_{\innv}} = \lg \half \lbr {\bf C}_{\innu},{\bf C}_{\innv} \rbr \rg - \lg {\bf C}_{\innu} \rg
\lg {\bf C}_{\innv} \rg , \quad \mathscr{C}_{\inc_{\innu} \ins_{\innv}} = \lg \half \lbr {\bf C}_{\innu},{\bf S}_{\innv} \rbr \rg - \lg 
{\bf C}_{\innu} \rg \lg {\bf S}_{\innv} \rg , \nn \\
\fl \quad \mathscr{C}_{\ins_{\innu} \inc_{\innv}} = \lg \half \lbr {\bf S}_{\innu},{\bf C}_{\innv} \rbr \rg - \lg {\bf S}_{\innu} \rg
\lg {\bf C}_{\innv} \rg , \quad \mathscr{C}_{\ins_{\innu} \ins_{\innv}} = \lg \half \lbr {\bf S}_{\innu},{\bf S}_{\innv} \rbr \rg - \lg 
{\bf S}_{\innu} \rg \lg {\bf S}_{\innv} \rg . \nn
\err
Next, we consider the discrete coherent states (\ref{e16}) as a reference state for all practical purposes here developed, as well as 
the mean values explicitly calculated in Table \ref{tab1}. In fact, let us foccus upon two closed-form results associated with the 
aforementioned RS uncertainty principles that will be used mainly in the body of the text. Thus, the first one consists in noticing that
$\mathscr{V}_{\innu} \mathscr{V}_{\innv}$ can be written in terms of a particular product involving the variances related to the sine
and cosine operators,
\be
\lb{b6}
\mathscr{V}_{\innu} \mathscr{V}_{\innv} = \lpar \mathscr{V}_{\inc_{\innu}} + \mathscr{V}_{\ins_{\innu}} \rpar \lpar \mathscr{V}_{\inc_{\innv}} + 
\mathscr{V}_{\ins_{\innv}} \rpar .
\ee
It is worth stressing that such result is valid for any quantum state properly defined in a finite-dimensional state vectors space. The next 
one refers to the constant uncertainty quantity ${\rm S}_{\mathscr{U}}$ evaluated through the sum of all inequalities (\ref{b2})--(\ref{b5}), namely,
\brr
\lb{b7}
\fl \qquad {\rm S}_{\mathscr{U}} &=& 1 - \mathscr{K}^{2}({\it D}_{\mathfrak{p}},0) - \mathscr{K}^{2}(0,{\it D}_{\mathfrak{q}}) -
\mathscr{K}^{2}({\it D}_{\mathfrak{p}},{\it D}_{\mathfrak{q}}) \nn \\
\fl & & + 2 \cos \lpar \frac{{\it D}_{\mathfrak{p}} {\it D}_{\mathfrak{q}}}{2 \hbar} \rpar \mathscr{K}({\it D}_{\mathfrak{p}},0) 
\mathscr{K}(0,{\it D}_{\mathfrak{q}}) \mathscr{K}({\it D}_{\mathfrak{p}},{\it D}_{\mathfrak{q}}) \geq 0 .
\err
Note that (\ref{b7}) coincides with the result obtained for the discrete vacuum state and does not depend on the discrete labels which characterize
the set of coherent states here studied. Indeed, ${\rm S}_{\mathscr{U}}$ depends only on the distances between successive eigenvalues of the discrete
position and momentum operators, as well as the dimension $N$ inherent to the states space once the ratio $\case{{\it D}_{\mathfrak{p}} 
{\it D}_{\mathfrak{q}}}{2 \hbar}$ is equivalent to $\case{\pi}{N}$ in such a case. In summary, this particular quantity describes the minimum bound
of uncertainty on the informational content related to the unitary operators ${\bf U}$ and ${\bf V}$.
\begin{table}[th]
\caption{\lb{tab1}Essential mean values used in the theoretical/numerical investigation of RS uncertainty principles related to the sine 
and cosine operators. The discrete coherent states established in this work lead us to obtain closed-form expressions which depend basically
on the discrete labels $\mathfrak{p}_{\kappa} \in \lbk -{\it R}_{\mathfrak{p}},{\it R}_{\mathfrak{p}} \rbk$ and $\mathfrak{q}_{\tau} \in 
\lbk -{\it R}_{\mathfrak{q}},{\it R}_{\mathfrak{q}} \rbk$, as well as the dimension $N$ of the states space.}
\begin{indented}
\item[] \begin{tabular}{@{}lllllllllllllllllllllll}
\br
{\sl Mean Values} & {\sl Discrete Coherent States} \\
\mr
$\lg {\bf C}_{\innu} \rg$ & $\cos \lpar \frac{{\it D}_{\mathfrak{p}} \mathfrak{q}_{\tau}}{\hbar} \rpar \mathscr{K}({\it D}_{\mathfrak{p}},0)$ 
\\ \\
$\lg {\bf S}_{\innu} \rg$ & $\sin \lpar \frac{{\it D}_{\mathfrak{p}} \mathfrak{q}_{\tau}}{\hbar} \rpar \mathscr{K}({\it D}_{\mathfrak{p}},0)$ 
\\ \\
$\lg {\bf C}_{\innv} \rg$ & $\cos \lpar \frac{{\it D}_{\mathfrak{q}} \mathfrak{p}_{\kappa}}{\hbar} \rpar \mathscr{K}(0,{\it D}_{\mathfrak{q}})$ 
\\ \\
$\lg {\bf S}_{\innv} \rg$ & $\sin \lpar \frac{{\it D}_{\mathfrak{q}} \mathfrak{p}_{\kappa}}{\hbar} \rpar \mathscr{K}(0,{\it D}_{\mathfrak{q}})$ 
\\ \\
$\lg {\bf C}_{\innu}^{2} \rg$ & $\half \lbk 1 + \cos \lpar \frac{2 {\it D}_{\mathfrak{p}} \mathfrak{q}_{\tau}}{\hbar} \rpar \mathscr{K}
(2 {\it D}_{\mathfrak{p}},0) \rbk$ \\ \\ 
$\lg {\bf S}_{\innu}^{2} \rg$ & $\half \lbk 1 - \cos \lpar \frac{2 {\it D}_{\mathfrak{p}} \mathfrak{q}_{\tau}}{\hbar} \rpar \mathscr{K}
(2 {\it D}_{\mathfrak{p}},0) \rbk$ \\ \\
$\lg {\bf C}_{\innv}^{2} \rg$ & $\half \lbk 1 + \cos \lpar \frac{2 {\it D}_{\mathfrak{q}} \mathfrak{p}_{\kappa}}{\hbar} \rpar \mathscr{K}
(0,2 {\it D}_{\mathfrak{q}}) \rbk$ \\ \\
$\lg {\bf S}_{\innv}^{2} \rg$ & $\half \lbk 1 - \cos \lpar \frac{2 {\it D}_{\mathfrak{q}} \mathfrak{p}_{\kappa}}{\hbar} \rpar \mathscr{K}
(0,2 {\it D}_{\mathfrak{q}}) \rbk$ \\ \\
$\lg \lbk {\bf C}_{\innu},{\bf C}_{\innv} \rbk \rg$ & $2 \rmi \sin \lpar \frac{{\it D}_{\mathfrak{p}} {\it D}_{\mathfrak{q}}}{2 \hbar} \rpar
\sin \lpar \frac{{\it D}_{\mathfrak{q}} \mathfrak{p}_{\kappa}}{\hbar} \rpar \sin \lpar \frac{{\it D}_{\mathfrak{p}} \mathfrak{q}_{\tau}}
{\hbar} \rpar \mathscr{K}({\it D}_{\mathfrak{p}},{\it D}_{\mathfrak{q}})$ \\ \\ 
$\lg \lbk {\bf C}_{\innu},{\bf S}_{\innv} \rbk \rg$ & $- 2 \rmi \sin \lpar \frac{{\it D}_{\mathfrak{p}} {\it D}_{\mathfrak{q}}}{2 \hbar} \rpar
\cos \lpar \frac{{\it D}_{\mathfrak{q}} \mathfrak{p}_{\kappa}}{\hbar} \rpar \sin \lpar \frac{{\it D}_{\mathfrak{p}} \mathfrak{q}_{\tau}}
{\hbar} \rpar \mathscr{K}({\it D}_{\mathfrak{p}},{\it D}_{\mathfrak{q}})$ \\ \\
$\lg \lbk {\bf S}_{\innu},{\bf C}_{\innv} \rbk \rg$ & $- 2 \rmi \sin \lpar \frac{{\it D}_{\mathfrak{p}} {\it D}_{\mathfrak{q}}}{2 \hbar} \rpar
\sin \lpar \frac{{\it D}_{\mathfrak{q}} \mathfrak{p}_{\kappa}}{\hbar} \rpar \cos \lpar \frac{{\it D}_{\mathfrak{p}} \mathfrak{q}_{\tau}}
{\hbar} \rpar \mathscr{K}({\it D}_{\mathfrak{p}},{\it D}_{\mathfrak{q}})$ \\ \\
$\lg \lbk {\bf S}_{\innu},{\bf S}_{\innv} \rbk \rg$ & $2 \rmi \sin \lpar \frac{{\it D}_{\mathfrak{p}} {\it D}_{\mathfrak{q}}}{2 \hbar} \rpar
\cos \lpar \frac{{\it D}_{\mathfrak{q}} \mathfrak{p}_{\kappa}}{\hbar} \rpar \cos \lpar \frac{{\it D}_{\mathfrak{p}} \mathfrak{q}_{\tau}}
{\hbar} \rpar \mathscr{K}({\it D}_{\mathfrak{p}},{\it D}_{\mathfrak{q}})$ \\
\br
\end{tabular}
\end{indented}
\end{table}

Now, let us adopt for convenience the discrete vacuum state as a reference guide in the numerical calculations. The choice of this particular 
quantum state permits us to verify that $\lg {\bf C}_{\innu} \rg$ and $\lg {\bf C}_{\innv} \rg$ are real and positive quantities, while the 
complementary mean values $\lg {\bf S}_{\innu} \rg$ and $\lg {\bf S}_{\innv} \rg$ are both null\footnote{These initial results have a counterpart
in those restrictions imposed by Massar and Spindel \cite{Massar} on the choice of phase for the unitary operators used in mathematical proof of 
the inequality
\bd
\left| \lg \Psi | \half \{ {\bf C}_{\innu},{\bf C}_{\innv} \} | \Psi \rg \right| \geq \sqrt{1 - \mathscr{V}_{\innu}} 
\sqrt{1 - \mathscr{V}_{\innv}} - \sqrt{\mathscr{V}_{\innu} \mathscr{V}_{\innv}} .
\ed
Despite certain guidelines adopted in this particular proof being not sufficiently clear, it is interesting to note that $\lg \mathfrak{q}_{\gamma} 
| \Psi \rg \equiv \lg \mathfrak{q}_{\gamma} | 0 \rg$ allows to establish a direct link with equation (\ref{e15}) for the odd $N$ case (even 
dimensionalities can also be dealt with simply by working on non-symmetrized intervals \cite{Ruzzi2}). Indeed, this perfect match leads us to
corroborate and generalize certain theoretical results previously obtained in \cite{Massar} for the discrete vacuum state.}. Furthermore, the 
RS uncertainty principles present certain mathematical peculiarities which deserve be mentioned. For instance, if one considers $N=3$ fixed, 
equations (\ref{b2}) and (\ref{b5}) are formally equivalent to the following algebraic identities:
\brr
\fl \quad \lbk \mathscr{K}(1,1) - \mathscr{K}(2,0) - 1 \rbk \lbk 4 \mathscr{K}^{2}(1,0) - \mathscr{K}(1,1) - \mathscr{K}(2,0) - 1 \rbk = 0 , \nn \\
\fl \quad \lbk 1 - \mathscr{K}(2,0) - \sqrt{3} \, \mathscr{K}(1,1) \rbk \lbk 1 - \mathscr{K}(2,0) + \sqrt{3} \, \mathscr{K}(1,1) \rbk = 0 . \nn
\err
These specific identities are completely satisfied for $\mathscr{K}(1,1) + \mathscr{K}(2,0) + 1 = 4 \mathscr{K}^{2}(1,0)$ and 
$1 - \mathscr{K}(2,0) = \sqrt{3} \, \mathscr{K}(1,1)$, both the equations being theoretically and numerically confirmed. In addition, let us also
mention that (\ref{b3}) and (\ref{b4}) are formally equivalent to inequality
\bd
\lbk 1 - \mathscr{K}(2,0) \rbk \lbk 1 + \mathscr{K}(2,0) - 2 \mathscr{K}^{2}(1,0) \rbk \geq 0
\ed
for any $N$ odd integer; namely, both the RS uncertainty principles are coincident in such a case. Briefly, let us comment on ${\rm S}_{\mathscr{U}}$ 
and its respective asymptotic limit: for $N \gg 1$, we verify immediately that ${\rm S}_{\mathscr{U}} \rightarrow 0$, which implies in a behaviour 
near to that observed for the continuous analogue (note that $\bi{Q}$ and $\bi{P}$ have continuous spectra in an infinite Hilbert space).

\begin{figure}[!t]
\centering
\begin{minipage}[b]{0.45\linewidth}
\includegraphics[width=\textwidth]{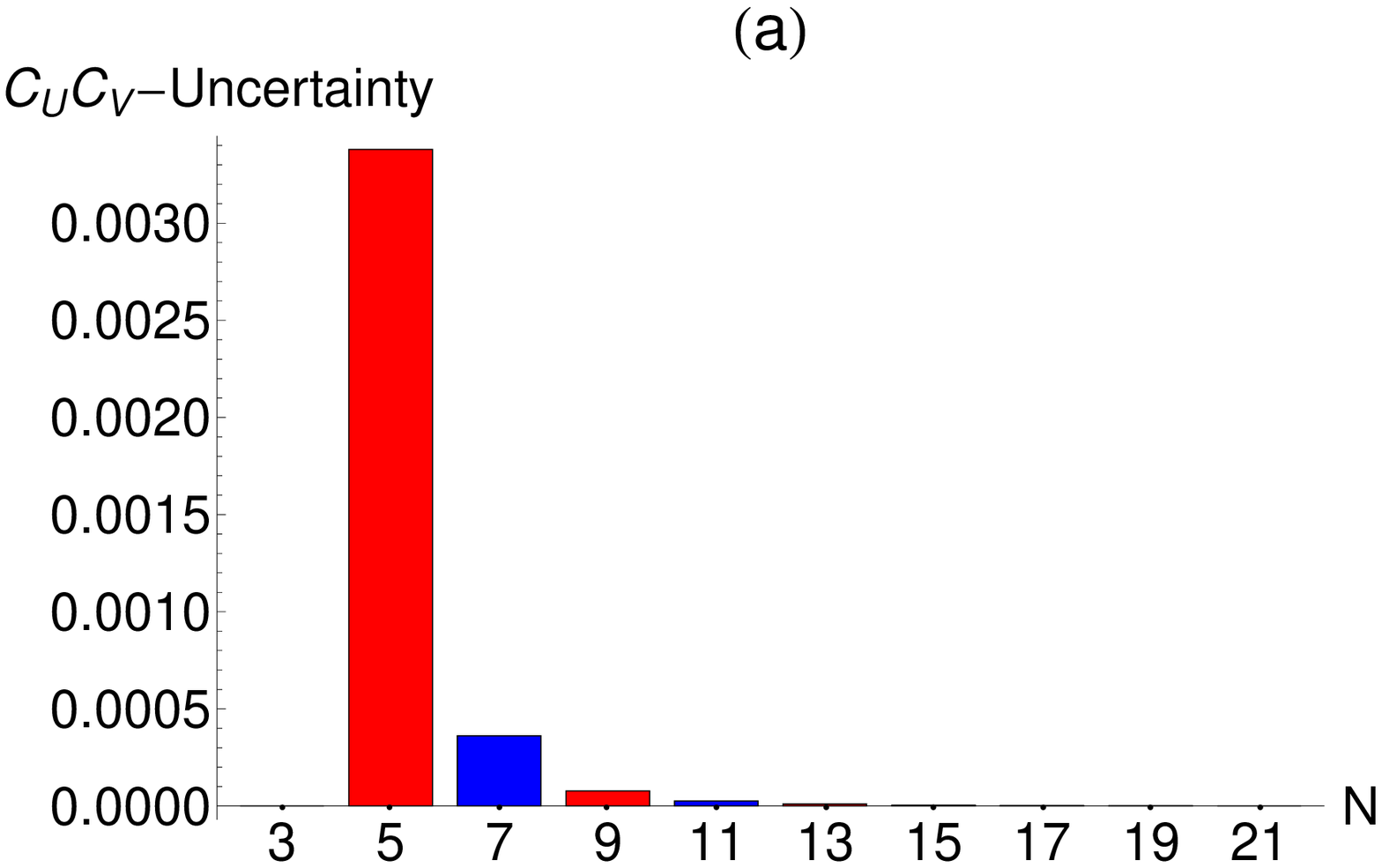}
\end{minipage} \hfill
\begin{minipage}[b]{0.45\linewidth}
\includegraphics[width=\textwidth]{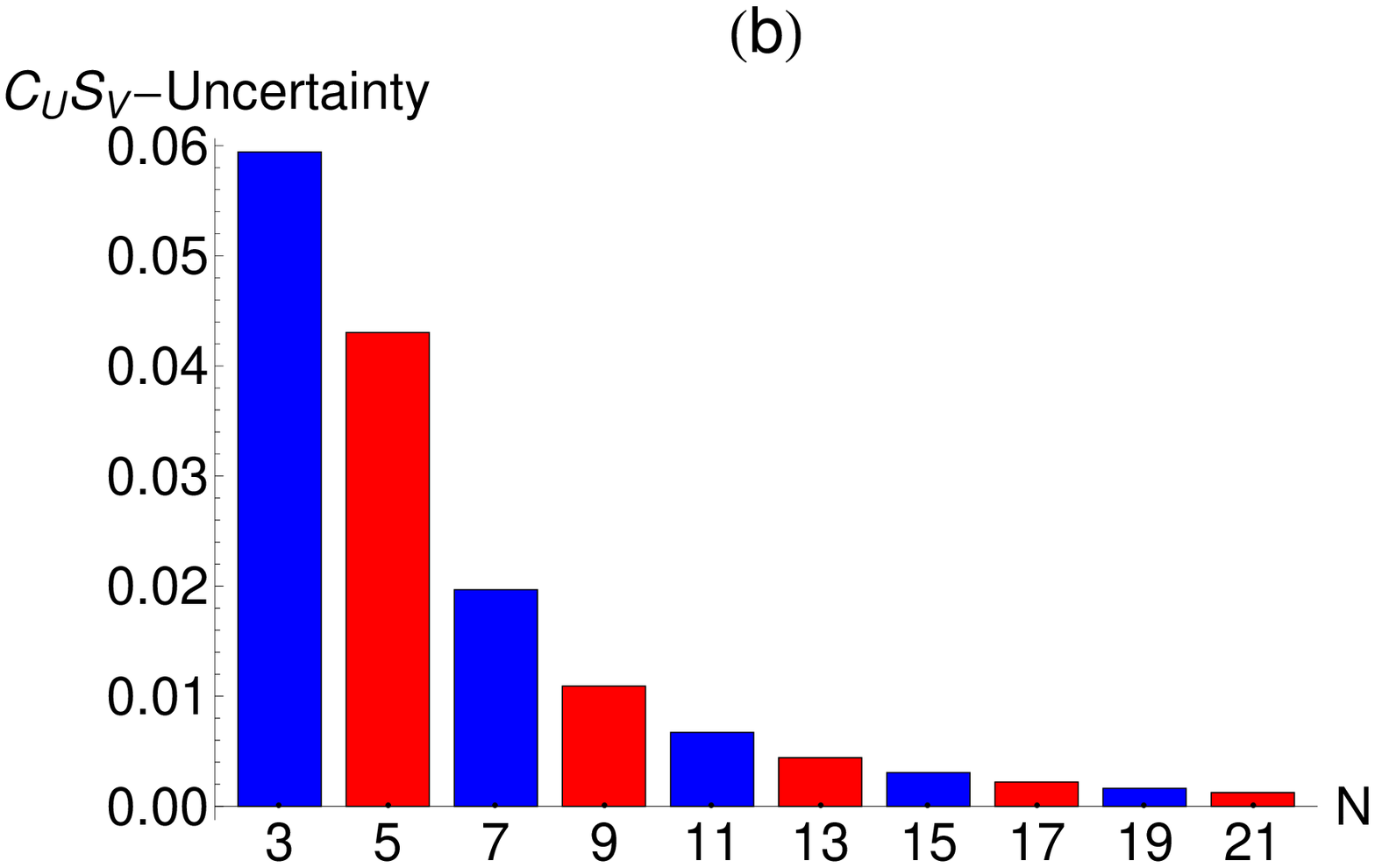}
\end{minipage} \hfill
\begin{minipage}[b]{0.45\linewidth}
\includegraphics[width=\textwidth]{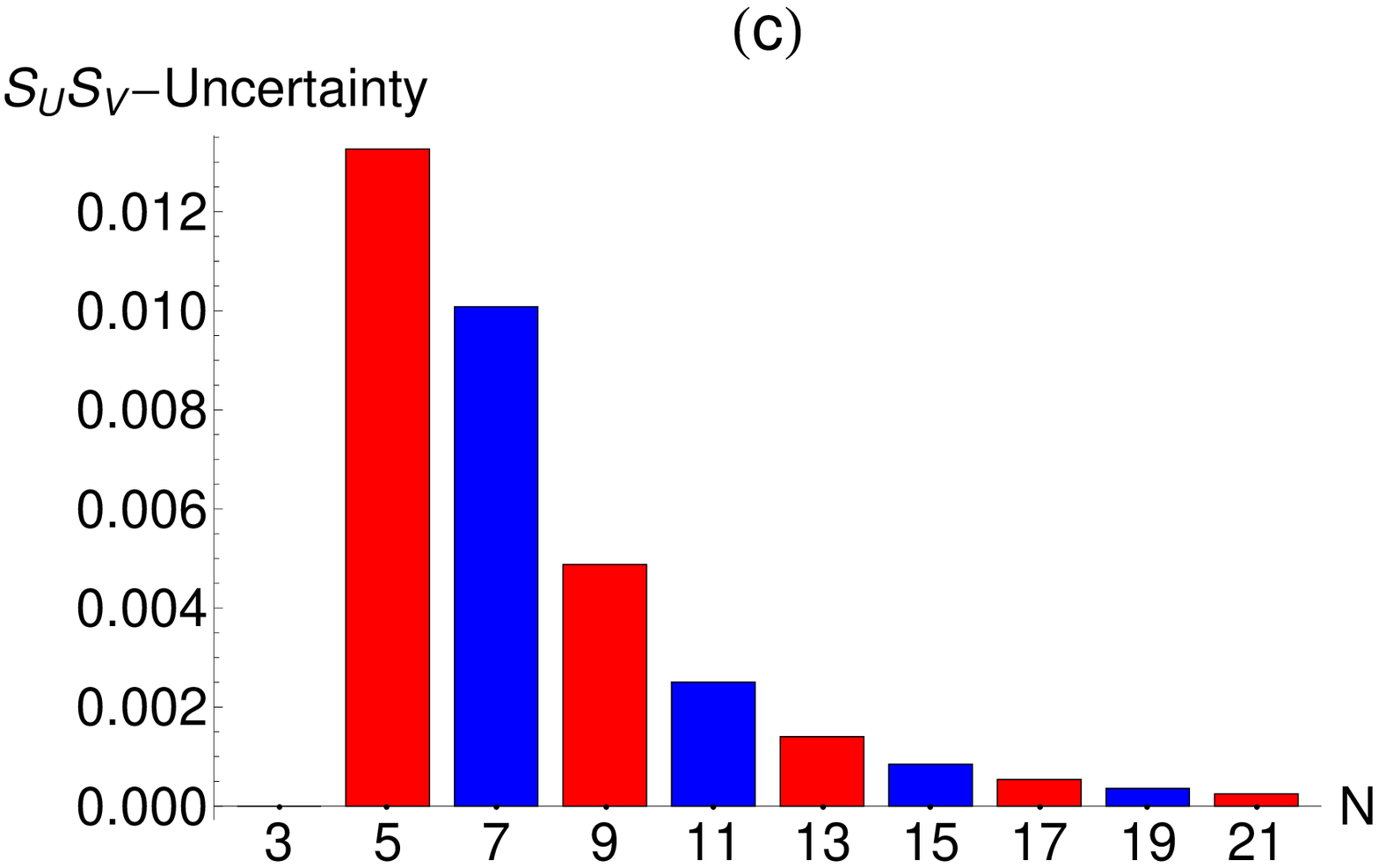}
\end{minipage} \hfill
\begin{minipage}[b]{0.45\linewidth}
\includegraphics[width=\textwidth]{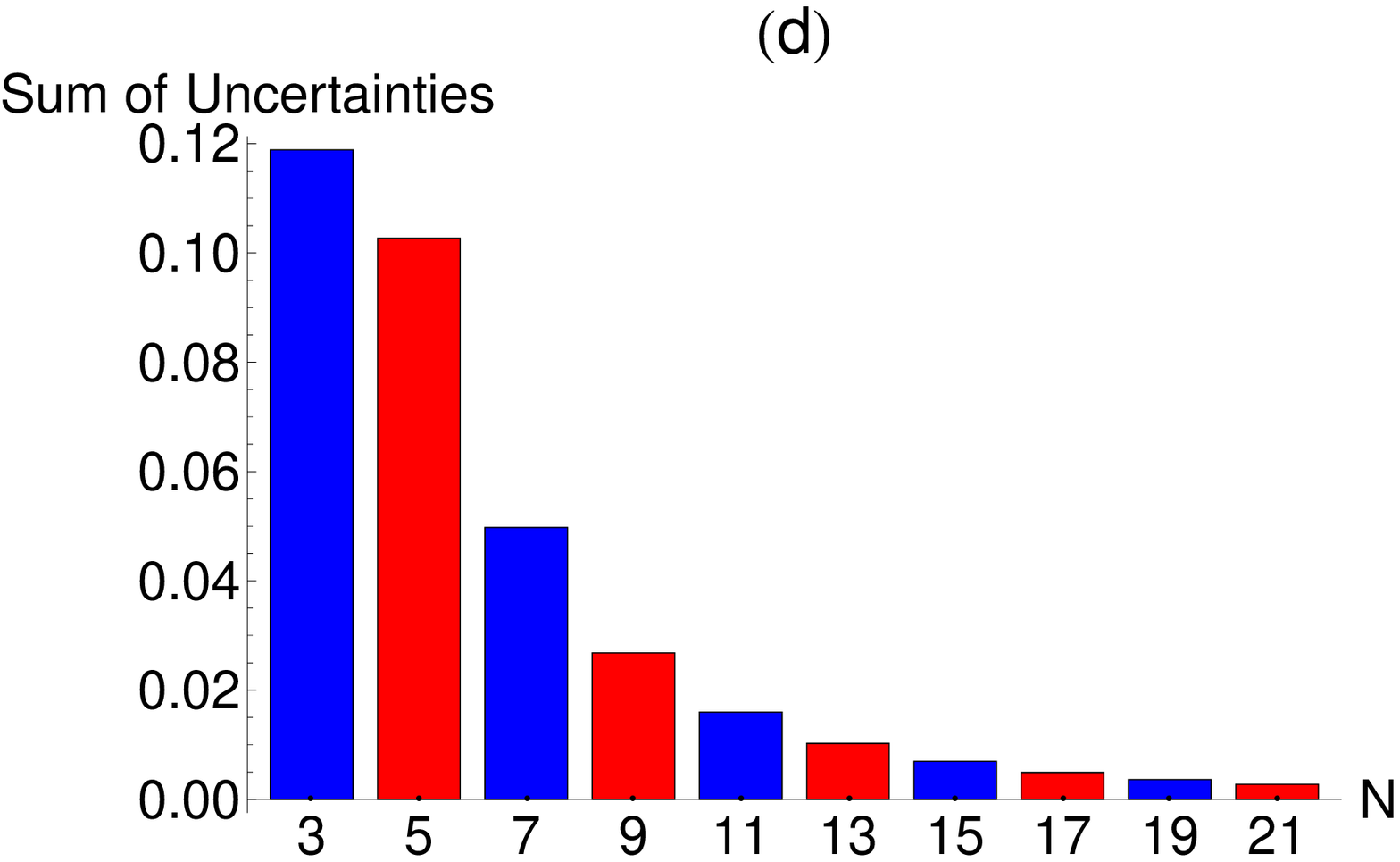}
\end{minipage}
\caption{Plots of $\mathscr{U}_{{\rm O}_{1} {\rm O}_{2}} - \case{1}{4} \left| \lg \lbk {\bf O}_{1},{\bf O}_{2} \rbk \rg \right|^{2}$ versus
$N$ for each relevant case of RS uncertainty principle associated with the sine and cosine operators previously discussed in this appendix.
It is important to stress that Table \ref{tab1} has a central role in the numerical calculations since it allows to adapt those analytical
results for the discrete vacuum state case. Furthermore, the asymptotic behaviour showed in these pictures indeed corroborate the continuous 
counterpart of the position and momentum operators evidenced in textbooks on quantum mechanics \cite{Jauch}.} 
\end{figure}
The next step then consists in providing graphics illustrations of such theoretical analysis. In this way, Figure B1 shows the plots of 
(a,b,c) $\mathscr{U}_{{\rm O}_{1} {\rm O}_{2}} - \case{1}{4} \left| \lg \lbk {\bf O}_{1},{\bf O}_{2} \rbk \rg \right|^{2}$ and (d)
${\rm S}_{\mathscr{U}}$ versus $N \in [3,21]$ for each relevant situation previously discussed. It is worth noticing that (a) and (c)
exhibit null values for $N=3$, while (b) reveals a significant contribution at the same situation. Moreover, (d) shows the total amount
of uncertainties associated with the measurement processes of certain physical properties inherent to the sine and cosine operators for 
the particular discrete vacuum state here studied. In summary, the results obtained from the numerical calculations corroborate, as expected,
the previous analysis about RS uncertainty principles; additionally, the formal results derived in this appendix also open new windows of future
investigations on the practical applications of this theoretical framework in different branches of physics\footnote{For technical details 
on the importance of the phase operators and their different definitions in the context of quantum mechanics and quantum optics see, for
example, Ref. \cite{Phase}.}. 


\end{document}